\documentclass[%
 reprint,
superscriptaddress,
 amsmath,amssymb,
pre,
]{revtex4-1}

\usepackage{graphicx}% Include figure files
\usepackage{dcolumn}% Align table columns on decimal point
\usepackage{bm}% bold math
\usepackage{color}
\usepackage[colorlinks=true,allcolors=blue,breaklinks=true]{hyperref}
\usepackage{physics}
\begin{document}

\preprint{APS/123-QED}

\title{Bridging particle deformability and collective response in soft solids}% Force line breaks with \\

\author{John D. Treado}
\thanks{These authors contributed equally.}
\affiliation{
 Department of Mechanical Engineering \& Materials Science, Yale University, New Haven, Connecticut 06520, USA
}
\affiliation{
 Integrated Graduate Program in Physical and Engineering Biology, Yale University, New Haven, Connecticut 06520, USA
}

\author{Dong Wang}
\thanks{These authors contributed equally.}
\affiliation{
 Department of Mechanical Engineering \& Materials Science, Yale University, New Haven, Connecticut 06520, USA
}

\author{Arman Boromand}
\affiliation{
 Department of Mechanical Engineering \& Materials Science, Yale University, New Haven, Connecticut 06520, USA
}

\author{Michael P. Murrell}
\affiliation{
 Department of Biomedical Engineering, Yale University, New Haven, Connecticut 06520, USA
}
\affiliation{
 Department of Physics, Yale University, New Haven, Connecticut 06520, USA
}
\affiliation{
 Systems Biology Institute, Yale University, West Haven, Connecticut 06516, USA
}

\author{Mark D. Shattuck}
\affiliation{
 Benjamin Levich Institute and Physics Department, The City College of New York, New York, New York 10031, USA
}

\author{Corey S. O'Hern}
\email{corey.ohern@yale.edu}
\affiliation{
 Department of Mechanical Engineering \& Materials Science, Yale University, New Haven, Connecticut 06520, USA
}
\affiliation{
 Department of Physics, Yale University, New Haven, Connecticut 06520, USA
}
\affiliation{
 Department of Applied Physics, Yale University, New Haven, Connecticut 06520, USA
}
\affiliation{
 Integrated Graduate Program in Physical and Engineering Biology, Yale University, New Haven, Connecticut 06520, USA
}

\date{\today}% It is always \today, today,

\begin{abstract}
Soft, amorphous solids such as tissues, foams, and emulsions are composed of deformable particles. However, the effect of single-particle deformability on the collective behavior of soft solids is still poorly understood. We perform numerical simulations of two-dimensional jammed packings of explicitly deformable particles to study the mechanical response of model soft solids. We find that jammed packings of deformable particles with excess shape degrees of freedom possess low-frequency quartic vibrational modes that stabilize the packings even though they possess fewer interparticle contacts than the nominal isostatic value. Adding intra-particle constraints can rigidify the particles, but these particles undergo a buckling transition and gain an effective shape degree of freedom when their preferred perimeter is above a threshold value. We find that the mechanical response of jammed packings of deformable particles with shape degrees of freedom differs significantly from that of jammed packings of rigid-shape particles, which emphasizes the importance of particle deformability in modelling soft solids.
\end{abstract}

\maketitle

\section{Introduction}

All soft, athermal solids deform in response to applied stress, yet much of our understanding of these systems relies on computational models using particles with fixed shapes~\citep{flow:DurianPRL1995,softp:vanHeckeJPCM2009}. While extensive work has focused on the effect of varying soft \emph{inter}particle interactions, less attention has been placed on how \emph{intra}particle degrees of freedom affect collective behavior. Foams~\citep{foams:BoltonPRL1990,hopper:BerthoPRE2006}, emulsions~\citep{emulsion:PrincenJColloidInterSci1983,jamming:BoromandSM2019}, and a wide array of living tissues~\citep{biofilm:SmithPNAS2017,biofilm:BerozNatPhys2018, collective:ParkNatMaterials2015,collective:TrepatNatPhys2018, shape:LecuitNatRevMolCellBio2007,actomyo:MurrellNatRevMolCellBio2015,cancer:JollyFrontOncol2015,dev:McMillenCurrBio2016,cancer:OswaldJPhysDAppPhys2017,dev:MongeraNature2018,cancer:IlinaNatCellBio2020,dev:KimBioRXiv2020, plants:WuytsPlantMethods2010,plants:KalveJExpBot2014,plants:SapalaELife2018,plants:MartinezThePlantCell2018,plants:BorsukBioRxiv2019} are composed of deformable objects. The complexity and variety of the shape degrees of freedom across these systems emphasizes the importance of investigating how single-particle deformability affects collective properties of soft solids, such as rigidity and linear response.

Athermal systems composed of soft particles form rigid solids at the jamming transition when all non-trivial deformations cost energy~\citep{jamming:OHernPRE2003}. If the particles are spherical, frictionless, and purely repulsive, it is well known that jamming occurs at an isostatic point; mechanically stable configurations at jamming onset in periodic boundary conditions with $N_{\rm dof}$ degrees of freedom and $N_c$ interparticle contacts satisfy $N_{\rm dof} - N_c = d - 1$~\citep{contacts:TkachenkoPRE1999}. This observation, a consequence of Maxwell-Calladine constraint counting~\citep{frames:PellegrinoIntJSolidsStructs}, has been used to rationalize the many anomalous mechanical and vibrational properties of jammed solids~\citep{jamming:SchreckPRL2011,softp:GoodrichNatPhys2014}. However, particles with non-spherical shapes typically jam with more degrees of freedom than interparticle contacts. These \emph{hypo}static packings gain mechanical stability from higher-order terms in the Taylor expansion of the potential energy~\citep{jamming:MailmanPRL2009,ellipse:DonevPRE2007,jamming:SchreckPRE2012,jamming:VanderWerfPRE2018,jamming:YuanSM2019}. Higher-order stability has been observed in jammed packings of a variety of non-spherical particles~\citep{jamming:VanderWerfPRE2018,jamming:YuanSM2019} and even in packings of ``breathing" particles that contain size degrees of freedom~\citep{nonspherical:BritoPNAS2018}. Higher-order constraints directly impact the vibrational spectrum~\citep{jamming:SchreckPRE2012,nonspherical:BritoPNAS2018}, shear response~\citep{jamming:SchreckPRL2011}, and the glass transition at finite temperature~\citep{jamming:ShenPRE2013}.

Recent work~\citep{rigidity:DamavindiArXiv2021} has proposed that such higher-order rigidity is a generic feature of hypostatic systems with sufficient pre-stress. This phenomenon has been used to explain the rigidity transition in Vertex models of confluent tissues~\citep{vertex:BiNatPhys2015,vertex:LePRX2019}, which can be viewed as dense packings of deformable polygonal cells that are constrained to be confluent. These results suggest that jammed packings of deformable particles might behave similarly, i.e. possess higher-order stability and mechanical and vibrational properties that diverge from those for jammed packings of frictionless, spherical particles. However, are jammed packings of deformable particles identical to those of non-spherical particles such as ellipses? Or does particle deformability lead to fundamentally different mechanical and vibrational response? And how do the properties of jammed packings change as the particles vary from highly deformable to completely rigid?  

\begin{figure}[t]
    \centering
    \includegraphics[width=0.5\textwidth]{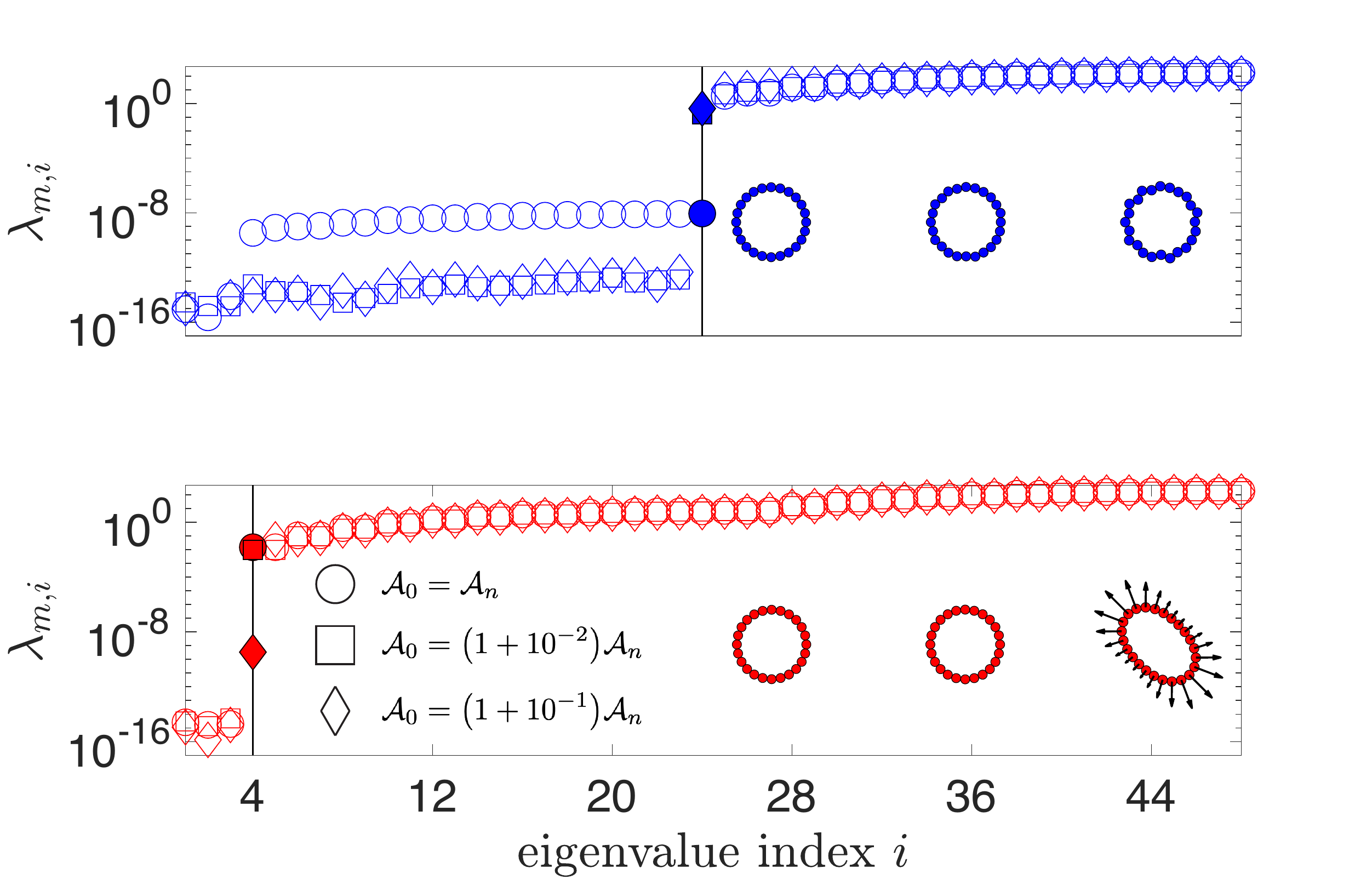}
\caption{\textbf{Single-particle vibrational spectra describe shape degrees of freedom.}  Eigenvalues of the single-particle dynamical matrix $\lambda_m$ for truly deformable particles (DP, top) and deformable particles with bending constraints (DPb, bottom) with $n = 24$ vertices. Symbols correspond to values of the preferred shape parameter $\mathcal{A}_0$, and $\mathcal{A}_n = n\tan(\pi/n)/\pi$ is the shape parameter of a regular $n$-gon. The vertical line at index $i = 24$ ($i = 4$) in the top (bottom) panel correspond to the crossover between zero and non-zero eigenvalues. Energy-minimized shapes are drawn in the insets, with $\mathcal{A}_0$ increasing from left to right, and the curvature vectors $\vec{\kappa}_i$ defined in Eq.~\eqref{eq:ub} are drawn on the buckled DPb particle. In both panels, $K_l = 1$, and $K_b = 10^{-2}$ in the bottom panel. }
    \label{fig:singleDP}
\end{figure}

In this article, we study the collective vibrational and mechanical properties of jammed solids composed of particles with varying degrees of deformability. In Sec.~\ref{sec:methods}, we introduce a model of deformable particles.  We define deformability through the single-particle vibrational spectra and show that the model can describe truly deformable and rigid-shape particles, as well as \emph{quasi}-deformable particles with characteristics between the two extremes. In Sec.~\ref{sec:results}, we investigate the rigidity, vibrational modes, and shear response in jammed solids composed of deformable particles. Our results emphasize that (a) packings of deformable particles in the rigid-shape-particle limit recover the properties found for jammed packings of soft spherical particles, but that (b) packings of truly deformable particles do not possess the same vibrational and mechanical properties as those for jammed packings of soft spherical particles. In Sec.~\ref{sec:conclusions}, we conclude with a discussion of the applicability of our results to glassy solids at finite temperature and to several experimental systems. We also include four appendices, which detail buckling in single particles with bending energy (Appendix~\ref{sec:appendix:buckling}), counting effective constraints using the dynamical matrix (Appendix~\ref{sec:appendix:constraints}), system-size dependence of the dynamical matrix and shear modulus (Appendix~\ref{sec:appendix:size}), and identification of the collective shape degrees of freedom (Appendix~\ref{sec:appendix:decomp}).

\section{Methods}~\label{sec:methods}
Systems of deformable particles in two dimensions are modeled by $N$ distinct polygons, each with $n_\mu$ vertices with positions $\vec{r}_{i\mu}$ for $i = 1,...,n_\mu$ and $\mu = 1,...,N$. Each polygon has an area $a_\mu$ and perimeter $p_\mu = \sum_{i=1}^{n_\mu} l_{i\mu}$, where $l_{i\mu}$ is the edge joining vertex $i$ and $i+1$ on polygon $\mu$. In previous work~\citep{jamming:BoromandPRL2018}, we studied the deformable polygon (DP) energy,
\begin{equation}\label{eq:dpEnergy}
    \begin{split}
        U_{\rm DP} &= \frac{\epsilon_a}{2}\sum_{\mu = 1}^{N} \qty(\frac{a_\mu}{a_{0\mu}} - 1)^2 + \frac{\epsilon_l}{2}\sum_{\mu = 1}^N\sum_{i=1}^{n_\mu} \qty(\frac{l_{i\mu}}{l_{0\mu}} - 1)^2 + U_{\rm int},
    \end{split}
\end{equation}
where $U_{\rm int}$ is the potential energy between interacting particles, and $\epsilon_a$ and $\epsilon_l$ are energies controlling area and perimeter fluctuations about the preferred areas $a_{0\mu}$ and edge lengths $l_{0\mu}$, respectively. Interactions between vertices $i$ and $j$ on cells $\mu$ and $\nu$ are governed by the pair potential $v$, which we assume depends only on the distance between two vertices, $r_{ij}^{\mu\nu} = \abs{\vec{r}_{i\mu} - \vec{r}_{j\nu}}$. We treat each vertex as a repulsive soft disk, where
\begin{equation}\label{eq:softDisks}
    v\qty(r_{ij}^{\mu\nu}) = \frac{\epsilon_c}{2}\qty(1 - \frac{r_{ij}^{\mu\nu}}{\sigma_{\mu\nu}})^2\Theta\qty(1 - \frac{r_{ij}^{\mu\nu}}{\sigma_{\mu\nu}}),
\end{equation}
$\sigma_{\mu\nu} = \qty(l_{0\mu} + l_{0\nu})/2$, each vertex has diameter $l_{0\mu}$, $\epsilon_c$ controls the strength of the interaction, and $\Theta$ is the Heaviside step function to enforce purely repulsive interactions. The total interaction energy is therefore $U_{\rm int} = \sum_{\nu, \mu} \sum_{i=1}^{n_{\mu}} \sum_{j=1}^{n_{\nu}} v\qty(r_{ij}^{\mu\nu})$, though we do not track overlaps between vertices $i$ and $i+1$ and $i$ and $i-1$ on the same particle. We measure lengths in units of the square root of the minimum preferred area, $\sqrt{\overline{a}_0}$, energies in units of $\epsilon_a$, and times in units of $\tau = \sqrt{\overline{a}_0/\epsilon_a}$, where all vertex masses have been set to $1$. The dimensionless \emph{preferred} shape parameter $\mathcal{A}_{0\mu} = \qty(n_\mu l_{0\mu})^2/(4\pi a_{0\mu})$ measures the amount of excess perimeter above a regular polygon with area $a_{0\mu}$ and thus controls particle deformability~\citep{jamming:BoromandPRL2018}. For the DP model, particle shapes depend only on $K_l = \epsilon_l/\epsilon_a$, $K_c = \epsilon_c/\epsilon_a$, and $\mathcal{A}_{0\mu}$. 

In Eq.~\eqref{eq:dpEnergy}, we see that the shape of a single DP particle is constrained by $n + 1$ terms given $n$ vertices, but each particle contains $2n$ degrees of freedom. By constraint counting, we expect $2n - (n+1) = n-1$ zero energy modes. While each particle contains two translational and one rotational degree of freedom that cannot be constrained by internal forces, DP particles still contain $n-4$ non-trivial zero modes. In this sense, DP particles are truly deformable and can change shape with zero energy cost. Example energy-minimized DP particles are shown in the top inset to Fig.~\ref{fig:singleDP}. 

To rigidify single DP particles, we add $n$ bending constraints along the particle perimeter~\citep{jamming:BoromandPRL2018},
\begin{equation}\label{eq:ub}
    U_{\rm b} = U_{\rm DP} + \frac{k_b}{2}\sum_{i=1}^n \vec{\kappa}_i^2,\hspace{0.3in} \vec{\kappa}_i = \frac{\vec{l}_i - \vec{l}_{i-1}}{l_0^2}.
\end{equation}
Eq.~\ref{eq:ub} has the additional parameter $K_b = k_b/(\epsilon_a l_0^2)$, which determines the energy cost of bending the particle perimeter. We refer to particles with this additional bending energy term as DPb particles.

In addition to single-particle properties, we also study configurations of multiple deformable particles near the onset of jamming. We prepare jammed packings of $N$ bidisperse ($50$:$50$ by number) deformable particles in square cells with side length $L$ and periodic boundary conditions. Small (large) particles are given $n_\mu = n_{\rm S}$ ($n_{\rm L}$) vertices with segment lengths $l_{0\mu}$ chosen such that $\mathcal{A}_{0\mu}/\mathcal{A}_n$ is identical for each particle, where $\mathcal{A}_n$ is the shape parameter of a regular $n$-gon. Therefore, when referring to the shape parameter chosen for a configuration of deformable particles, we will use $\mathcal{A}_0$ to mean the ratio of $\mathcal{A}_0/\mathcal{A}_n$ for a particle with a given number of vertices. We choose $n_{\rm L}$ to be the nearest integer to $1.4n_{\rm S}$ to enforce an approximate $1.4$ large-to-small size ratio to avoid crystallization and phase separation~\citep{jamming:ZhangPRE2014}. Likewise, large particles are given preferred areas $a_{0\mu} = (1.4)^2\overline{a}_0$. To create jammed packings, we randomly place particles in the simulation cell at low packing fraction $\phi$ and isotropically compress the system by increasing the particle size. Compression steps are followed by minimization of the total potential energy $U$ using FIRE. We take configurations as sufficiently near an energy minimum when the total root-mean-square force $< 10^{-12}$. We monitor jamming onset using the virial pressure $P = (\Sigma_{xx} + \Sigma_{yy})/2$, where the virial stress is
\begin{equation}\label{eq:virialStress}
    \Sigma_{\xi\xi'} = \epsilon_c L^{-2}\sum_{\nu\neq\mu}\sum_{i=1}^{n_\mu}\sum_{j=1}^{n_\nu}\qty(1 - \frac{r_{ij}^{\mu\nu}}{\sigma_{ij}^{\mu\nu}})\frac{r_{ij,\xi}^{\mu\nu}r_{ij,\xi'}^{\mu\nu}}{r_{ij}^{\mu\nu}\sigma_{ij}^{\mu\nu}}.
\end{equation}
$r_{ij,\xi}^{\mu\nu}$ is the $\xi$-component of the vector separating vertex $i$ on cell $\mu$ and vertex $j$ on cell $\nu$ and $\xi = x$ or $y$. We identify jamming onset, with packing fraction $\phi_J$, when the pressure $10^{-7} < P < 2 \times 10^{-7}$. We have confirmed that the results presented below do not depend on the pressure threshold as long as it is sufficiently small. Throughout this work, we will use $K_l = K_c = 1$ unless otherwise stated. 

\section{Results}~\label{sec:results}
\subsection{Rigidity}~\label{sec:results:rigidity}
We first investigate the rigidity of single DP and DPb particles by normal mode analysis. Single-particle normal modes are eigenvectors of the dynamical matrix $\mathcal{M}$, with block elements defined by $\mathcal{M}_{ij} = \pdv*{U}{\vec{r}_i}{\vec{r}_j}$. In Fig.~\ref{fig:singleDP}, we plot the normal mode eigenvalues $\lambda_m$ for DP and DPb particles with $n = 24$ vertices and varying preferred shape parameters $\mathcal{A}_0$. We find DP particles have $n-1$ near-zero modes ($\lesssim 10^{-15}$) when $\mathcal{A}_0 > 1$, as expected from constraint counting. Interestingly, the DP particle with $\mathcal{A}_0 = 1$ possess $n-3$ \emph{low} frequency modes significantly above the noise floor. Although the particle shape is underconstrained, particles with $\mathcal{A}_0 = 1$ cannot deform without increasing their perimeter-to-area ratio. These DP particles therefore are stabilized by prestress, a phenomenon found in underconstrained tensegrity structures~\citep{frames:PellegrinoIntJSolidsStructs} and disordered spring networks~\citep{rigidity:DamavindiArXiv2021}. 

\begin{figure}
    \centering
    \includegraphics[width=0.35\textwidth]{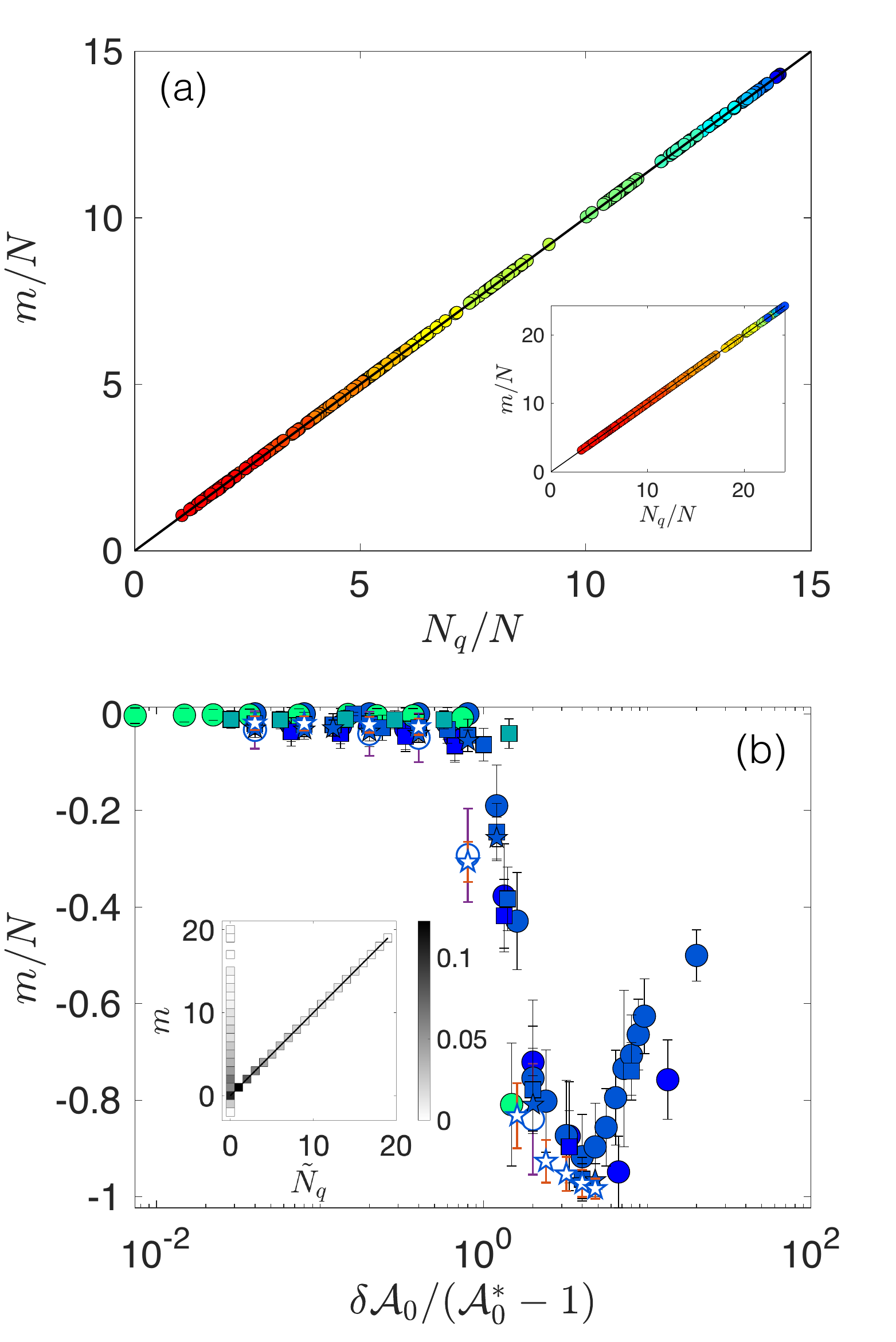}
    \caption{\textbf{Deformable particles do not typically jam at isostaticity.} (a) Number of missing contacts per particle $m/N$ in packings of $N = 64$ DP particles with $n_{\rm S} = 16$ (inset, $n_{\rm S} = 24$) vs. number of quartic modes per particle $N_q/N$. Black solid line gives $m = N_q$, and colors represent shape parameter values from $\mathcal{A}_0 = 1.0001$ to $1.24$, sorted from low (blue) to high (red) values. (b) Missing contacts per particle $m/N$, where now $m = 3N' - 1 - N_{\rm vv}$ for a system with $N'$ non-rattler particles, in packings of DPb particles plotted vs. $\delta \mathcal{A}_0/\qty(\mathcal{A}^*_0 - 1)$. $\mathcal{A}_0^*$ is the particular buckling shape parameter for a given set of parameters, and $\delta\mathcal{A}_0 = \mathcal{A}_0 - 1$. Colors represent $K_b$ (sorted from blue to green), spanning $K_b = 0.005$ to $0.05$. The filled symbols are for $n_{\rm S} = 16$, while white symbols are for $n_{\rm S} = 24$, and shapes represent different system sizes: $N = 16$ (circles), $32$ (squares), and $64$ (stars). The inset shows $m = 4N' - 1 - N_{\rm vv}$ for jammed packings of $N = 64$ buckled DPb particles with $N'$ non-rattler particles, and $\tilde{N}_q$ is the number of apparent higher-order modes inferred by the heuristic counting described in the main text. Shape parameters from $\mathcal{A}_0 = 1.04$ to $1.12$ are shown, and darker color signifies probability density as denoted by the colorbar. The black line gives $m = \tilde{N}_q$.}
    \label{fig:isostaticity}
\end{figure}

As DPb particles contain $n$ additional constraints, we expect them to behave as rigid-shape particles (such as frictionless soft disks or ellipses) where any shape deformation costs energy. In Fig.~\ref{fig:singleDP}, we find that there are only $3$ near-zero modes, corresponding to the trivial zero modes, for DPb particles with sufficiently small preferred shape parameter and that the particles energy-minimize to regular polygons. However, when $\mathcal{A}_0 = 1.1$, the DP particle is ``buckled" with an elliptical shape and an additional low-frequency mode $\lambda_{m,4} \approx 10^{-10}$.

In Appendix~\ref{sec:appendix:buckling}, we show that the DPb model contains a buckling transition where energy-minimized shapes elongate from regular polygons and the first non-trivial normal mode eigenvalue $\lambda_{m,4}$ drops from $\sim K_b$ to near zero. The transition point, $\mathcal{A}_0^*$, varies for different $K_b$, but the behavior after buckling is similar: $\lambda_{m,4}$ rises from the noise floor with increasing $\mathcal{A}_0$, and particles increasingly elongate. The small value of $\lambda_{m,4}$ after buckling suggests that buckled DPb particles gain an extra degree of freedom even though the number of constraints remains constant, a feature reminiscent of the rigidity transition in vertex models of confluent tissues~\citep{vertex:BiNatPhys2015,rigidity:DamavindiArXiv2021} and topological metamaterials~\citep{metamat:LubenskyNatPhys2014,metamat:ChenPNAS2014}.

We then investigate rigidity in jammed packings of DP and DPb particles by calculating the collective vibrational response. In a jammed packing of $N'$ non-rattler DP particles with $\overline{n}$ vertices per particle on average, there are $2N'\overline{n}$ degrees of freedom, $N'(\overline{n}+1)$ shape constraints and $N_{\rm vv}$ vertex-vertex contacts to constrain the shape degrees of freedom. Isostaticity would dictate $N_{\rm vv} = N'\qty(\overline{n}-1) - 1$, but in Fig.~\ref{fig:isostaticity} (a) we show that DP particles at jamming onset are \emph{hypo}static and seemingly missing the requisite number of interparticle contacts for jamming. The number of missing contacts for jammed DP particles is $m = N'\qty(\overline{n}-1) - 1 - N_{\rm vv}$.

Hypostaticity at jamming onset is often observed in packings of non-spherical particles~\citep{jamming:VanderWerfPRE2018}. Recent work has shown that these systems are stabilized by higher-order \emph{quartic} modes of the potential energy~\citep{jamming:MailmanPRL2009,jamming:VanderWerfPRE2018,nonspherical:BritoPNAS2018}. In Appendix~\ref{sec:appendix:constraints}, we show that quartic modes can be identified by decomposing the dynamical matrix $\mathcal{M}$ into the stiffness $\mathcal{H}$ and stress $\mathcal{S}$ matrices~\citep{ellipse:DonevPRE2007,jamming:SchreckPRE2012}. As shown in Fig.~\ref{fig:isostaticity} (a), we find that the number of missing contacts in jammed DP packings always matches the number of quartic modes $N_q$ across a wide range of shape parameters. 

\begin{figure}
    \centering
    \includegraphics[width=0.5\textwidth]{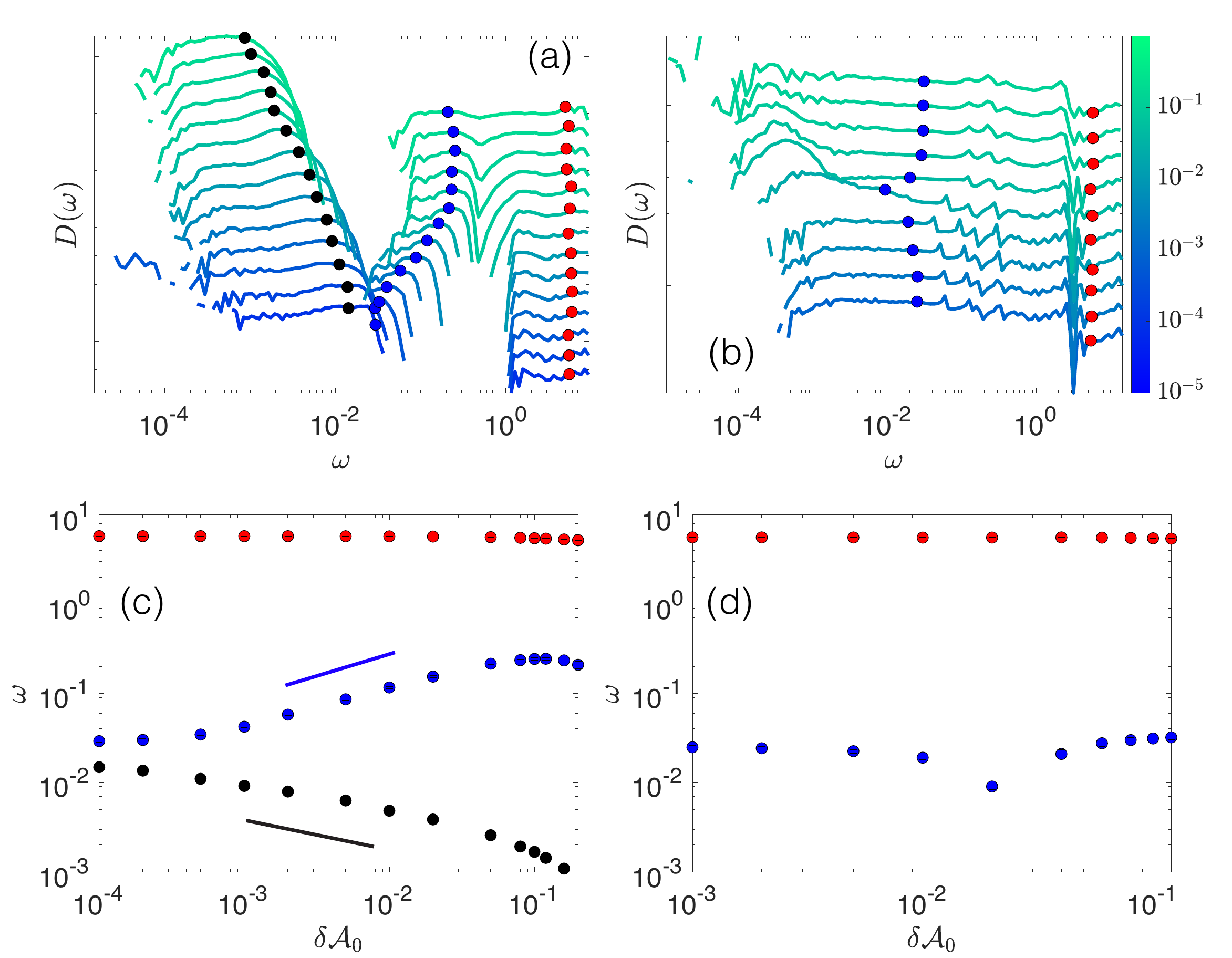}
    \caption{\textbf{Quartic modes and buckling strongly influence the low-frequency behavior of the vibrational response.} Vibrational density of states $D(\omega)$ at jamming onset for the DP and DPb models and a range of shape parameters. (a) $D(\omega)$ for $N = 64$ jammed, bidisperse DP particles ($n_{\rm S} = 16$) with shape parameter $1.0001 \leq \mathcal{A}_0 \leq 1.20$. (b) $D(\omega)$ for jammed packings of DPb particles ($n_{\rm S} = 24$) with $K_b = 10^{-2}$ and shape parameters $1.001 \leq \mathcal{A}_0 \leq 1.12$. In both (a) and (b), curves are offset for clarity, the perimeter spring constant $K_l = 1$, and the curve colors shown in the colorbar represent $\delta \mathcal{A}_0 = \mathcal{A}_0 - 1$. (c), (d) Characteristic frequencies $\omega_0$ (black), $\omega_1$ (blue), and $\omega_2$ (red) as a function of $\delta A_0$ for DP (left) and DPb particles (right). The black and blue lines in (c) represent the scalings $\sim\delta\mathcal{A}_0^{-1/3}$ and $\sim\delta\mathcal{A}_0^{1/2}$, respectively. Dots in (a) and (b) represent the location of each characteristic frequency in $D(\omega)$.}
    \label{fig:vdos}
\end{figure}

As DPb particles with $n$ vertices contain $n$ additional constraints, we expect them to behave similarly to packings of rigid-shape bumpy particles that are known to be isostatic at jamming~\citep{jamming:PapaPRL2013}. Indeed, when the particles are regular polygons, i.e. $\mathcal{A}_0 < \mathcal{A}_0^*$, we show in Fig.~\ref{fig:isostaticity} (b) that these packings are isostatic and the number of contacts $N_{\rm vv}$ equals $3N' - 1$, the total number of contacts expected for an isostatic packing of $N'$ non-rattler particles, each with $3$ degrees of freedom. However, in Fig.~\ref{fig:isostaticity} (b) we show that near the buckling transition $\mathcal{A}_0^*$, packings gain contacts and appear to be \emph{hyper}static at jamming onset. Hyperstaticity at jamming onset is extremely rare when using athermal protocols~\citep{jamming:SchreckPRE2011,jamming:TuckmanSM2020}, so it seems the ``buckling mode" with low eigenvalue effectively gives the DPb particles an extra degree of freedom, making these packings actually \emph{hypo}static with $N_{\rm vv} < 4N' - 1$. 

One might expect to be able to count missing contacts for packings of DPb particles by decomposing the dynamical matrix $\mathcal{M}$ into $\mathcal{H}$ and $\mathcal{S}$ as we did for packings of DP particles. However, we show in Appendix~\ref{sec:appendix:constraints} that all non-trivial eigenvalues of the stiffness matrix $\mathcal{H}$ are nonzero for DPb packings and several orders of magnitude larger than the eigenvalues of the dynamical matrix $\mathcal{M}$. As discussed in Ref.~\citep{rigidity:DamavindiArXiv2021}, the presence of positive eigenvalues of the stress matrix $\mathcal{S}$ make counting missing constraints via the dynamical matrix indeterminate. Despite this, we find some evidence of missing contacts in DPb packings using a heuristic approach detailed in Appendix~\ref{sec:appendix:constraints}. Briefly, if a packing of DPb particles is missing $m = 4N' - 1 - N_{\rm vv}$ contacts, we check (i) whether there is a gap between the $m$th non-trivial eigenvalue $\lambda_m$ of $\mathcal{M}$ and $\lambda_{m+1}$, or (ii) if there is no apparent gap, wherether the $m$th non-trivial mode has a significantly larger participation ratio than mode $m+1$. We show in the inset of Fig.~\ref{fig:isostaticity} (b) that the heuristic counting largely identifies correctly the $\tilde{N}_q$ higher-order modes that stabilize the missing contacts. However, there are several cases where the counting fails, highlighting the difficulty in determining rigidity in packings of DPb particles with negative pre-stress. Notably, most cases in which the correct number of missing contacts could not be identified in the dynamical matrix eigenvalue spectra or mode structure (i.e., when $m \neq \tilde{N}_q$) occur at $\tilde{N}_q = 0$. That is, whenever there is a notable gap or change in eigenmode participation ratio, we correctly count the number of missing contacts. We reserve a more in-depth analysis of the edge cases where missing contacts were not identified by $\mathcal{M}$, as well as a predictive theory for the missing contacts as a function of shape parameter, for future work.

\subsection{Vibrational response}~\label{sec:results:vdos}
\begin{figure}
    \centering
    \includegraphics[width=0.4\textwidth]{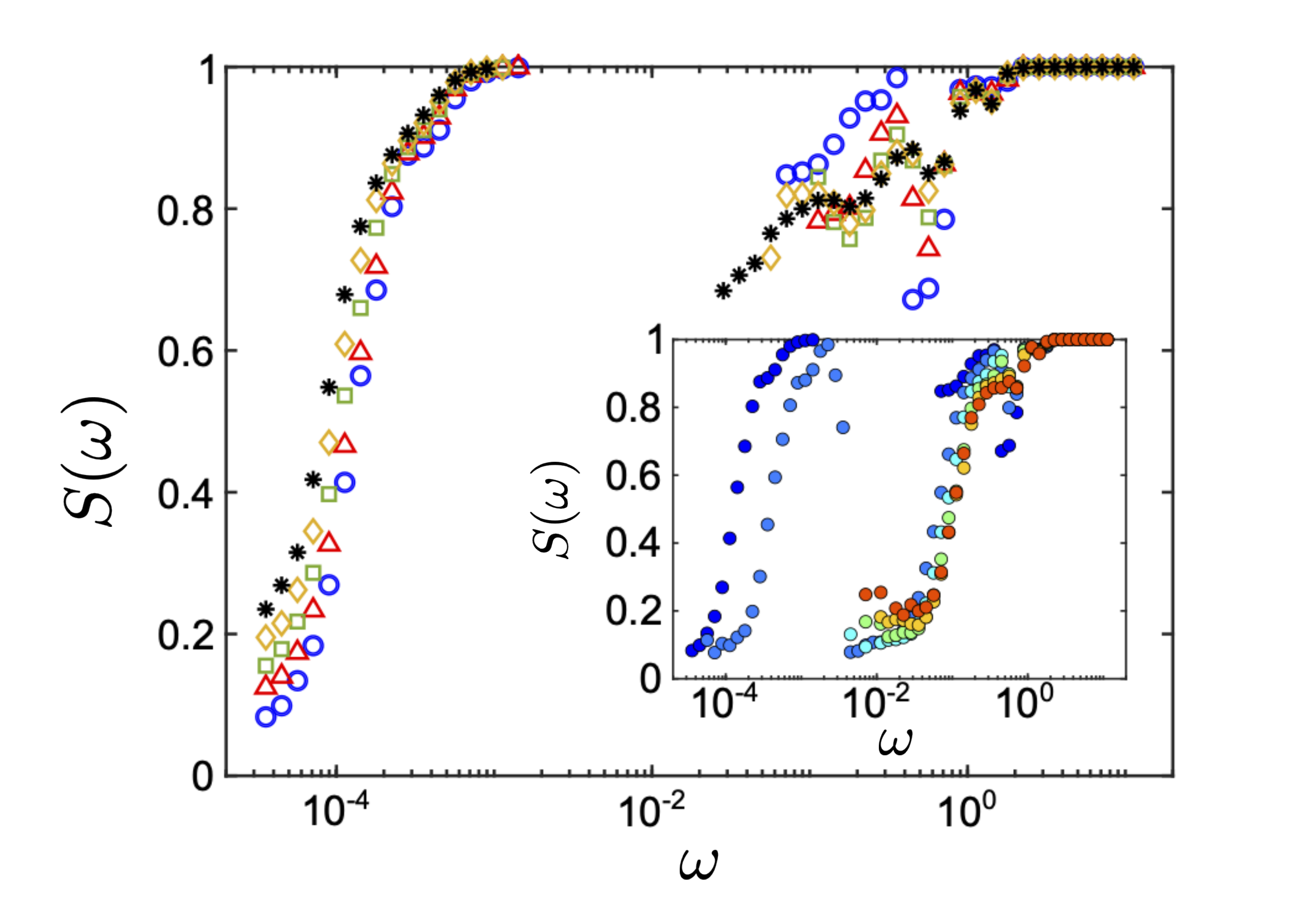}
    \caption{\textbf{Collective shape degrees of freedom are important in the vibrational response of jammed DP particles.} Magnitude of mode projection onto the shape degrees of freedom ($S$) versus the eigenmode frequency $\omega$ of the dynamical matrix for $N = 256$ DP packings with $n_{\rm S} = 16$ and $\mathcal{A}_0 = 1.02$ (circles), $1.06$ (triangles), $1.1$ (squares), $1.14$ (diamonds), and $1.18$ (asterisks). Inset: $S(\omega)$ for DP packings with $\mathcal{A}_0 = 1.02$ at several packing fractions from $\phi = \phi_J$ (blue circles) to $0.98$ (red circles) in increments of $2 \times 10^{-2}$.}
    \label{fig:modeProjection}
\end{figure}

We next study the density of vibrational states $D(\omega)$ for non-trivial vibrational modes with frequency $\omega_i = \sqrt{\lambda_{m,i}}$. In packings of DP particles, we observe three distinct bands of vibrational response in Fig.~\ref{fig:vdos} (a) due to quartic modes (with mean frequency $\omega_0$), mid-frequency collective modes (consistently the first $N-1$ quadratic modes, with mean frequency $\omega_1$), and high-frequency shape modes (with mean frequency $\omega_2$). As shown in Appendix~\ref{sec:appendix:size}, $D(\omega)$ and the characteristic frequencies do not vary significantly with system size. A similar three-band structure is found in the vibrational response of jammed packings of rigid-shape non-spherical particles~\citep{jamming:VanderWerfPRE2018}, although for the DP packings, the second band of modes corresponds to shape fluctuations at particle-particle interfaces rather than particle rotations. Additionally, as shown in Fig~\ref{fig:vdos} (c), we find the characteristic scaling $\omega_0 \sim \delta \mathcal{A}_0^{-1/3}$, indicating collective motion becomes less costly as particles become more deformable. We note this behavior differs from jamming of frictionless non-spherical particles with rigid shape~\citep{jamming:SchreckPRE2012,nonspherical:BritoPNAS2018}, where $\omega_0\sim \mathcal{A}_0^{1/2}$. We find approximate $1/2$ scaling with shape parameter in the mid-frequency band $\omega_1$, although this exponent is $\sim 1$ in packings of rigid-shape non-spherical particles. The stiff shape mode band with mean frequency $\omega_2$ does not vary with particle shape. 

Previous studies have argued that driven and jammed amorphous solids can be described using spherical particles with soft interparticle potentials~\citep{flow:DurianPRL1995}, where particle deformability is modelled by large interparticle overlaps. However, the deviation in $D(\omega)$ for packings of DP particles from that for soft non-spherical particles suggests that explicit shape change plays an important role in determining the vibrational response of soft particles. We further investigate the effect of shape change in the vibrational response by computing the projection of each eigenmode onto collective particle translations ($T$), rotations ($S$), and shape degrees of freedom ($S$), as described in Appendix~\ref{sec:appendix:decomp}. In Fig.~\ref{fig:modeProjection}, we show that even the low-frequency modes of jammed DP particles have a significant collective shape projection $S$ across a wide range of shape parameters ($1.02 \leq \mathcal{A}_0 \leq 1.18$). We find that $S(\omega)$ remains $> 0$ at the lowest frequencies even when the compression is increased well above jamming onset.  Explicit shape change is therefore necessary to capture important features of driven soft materials, such as flows of bubbles~\citep{hopper:BerthoPRE2006}, droplets \cite{droplets:FoglinoPRL2017} and emulsions~\citep{weeks:HongPRE2017,emulsions:GolovkovaSM2020}.

\begin{figure}
    \centering
    \includegraphics[width=0.5\textwidth]{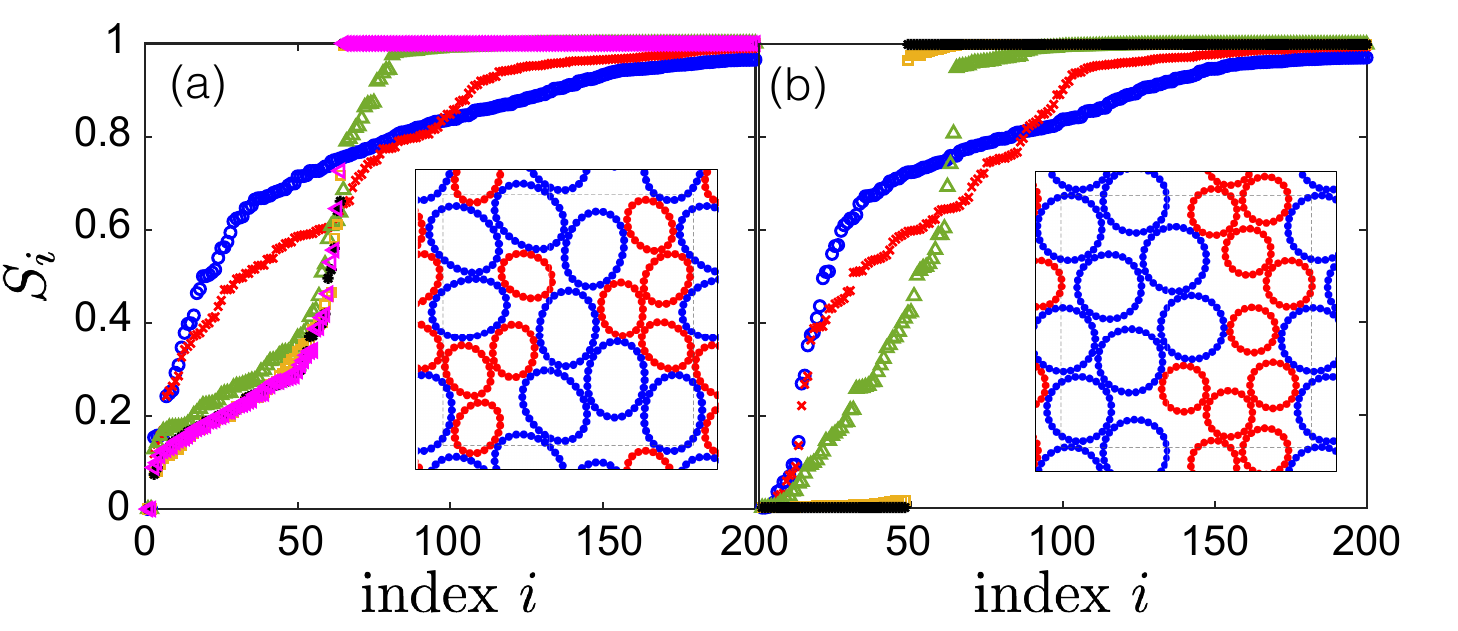}
    \caption{\textbf{The rigid-shape-particle limit exists for packings of DPb particles when $\mathcal{A}_0 < \mathcal{A}_0^*$.} (a) Shape degrees of freedom per eigenmode, sorted from smallest to largest, in a packing of $N = 16$ DPb particles (shown in inset) with $n_{\rm S} = 24$, $K_l = 1$, $K_b = 10^{-2}$, and $\mathcal{A}_0 = 1.04$, which is above the buckling transition $\mathcal{A}_0^* = 1.03$. Curves show changing the interaction parameter $K_c$ from $K_c = 1$ (blue circles) to $K_c = 10^{-5}$ (magenta left triangles) with intermediate values spaced by a factor of $10$. (b) Same as (a), but now $\mathcal{A}_0 = 1$ and particles are regular polygons when energy minimized (as shown in inset). $K_c$ is now varied from $K_c = 1$ (blue circles) to $K_c = 10^{-4}$ (black asterisks). The symbols for intermediate values of $K_c$ are the same as in (a).}
    \label{fig:rigidLimit}
\end{figure}

We also computed $D(\omega)$ for jammed DPb particles as shown in Fig.~\ref{fig:vdos} (b). These systems no longer have a distinct band structure in $D(\omega)$, as there are no obvious quartic modes. Here, we define $\omega_1$ as the mean of the first $N-1$ modes after the trivial zeros in analogy with the DP packings, and $\omega_2$ is the mean of all other modes. For systems with $\mathcal{A}_0 < \mathcal{A}_0^*$, $D(\omega)$ is relatively unchanged as a function of $\mathcal{A}_0$. For packings with buckled DPb particles  ($\mathcal{A}_0 > \mathcal{A}_0^*$), we observe a higher density of low-frequency modes near the buckling transition and a cusp in $\omega_1$ at $\mathcal{A}_0^*$ as shown in  Fig.~\ref{fig:vdos} (d). The abundance of low-frequency modes is likely due to the sudden decrease in the magnitude of the single-particle $\lambda_{m,4}$ mode at the buckling transition (see Appendix~\ref{sec:appendix:buckling}), and the appearance of modes that can stabilize more than one degree of freedom. 

The large density of low frequency modes at the buckling transition for DPb particles raises an important question. Is there a regime where DPb particles will behave as particles with rigid shapes? Or are DPb particles \emph{quasi}-deformable with persistent non-rigid-shape behavior? To address this question, we compute the collective shape degrees of freedom $S$ in individual jammed packings of DPb particles in the rigid-shape limit ($K_c \to 0$). We show in Fig.~\ref{fig:rigidLimit} (b) that non-buckled DPb particles ($\mathcal{A}_0 < \mathcal{A}_0^*$) eventually reach the rigid-shape limit ($K_c \lesssim 10^{-4}$), where the first $3N$ eigenmodes correspond to purely translational and rotational degrees of freedom and the rest of the spectrum contains only shape degrees of freedom. However, when particles buckle (i.e. $\mathcal{A} > \mathcal{A}_0^*$ in Fig.~\ref{fig:rigidLimit} (a)), $S$ is non-zero for the first $3N$ eigenmodes for all values of $K_c$. We conclude that DPb particles that remain regular polygons are effectively rigid-shape particles, whereas buckled DPb particles are quasi-deformable, and thus the shape degrees of freedom play a key role in their vibrational response.

\subsection{Shear response}~\label{sec:results:shear}
\begin{figure}
    \centering
    \includegraphics[width=0.45\textwidth]{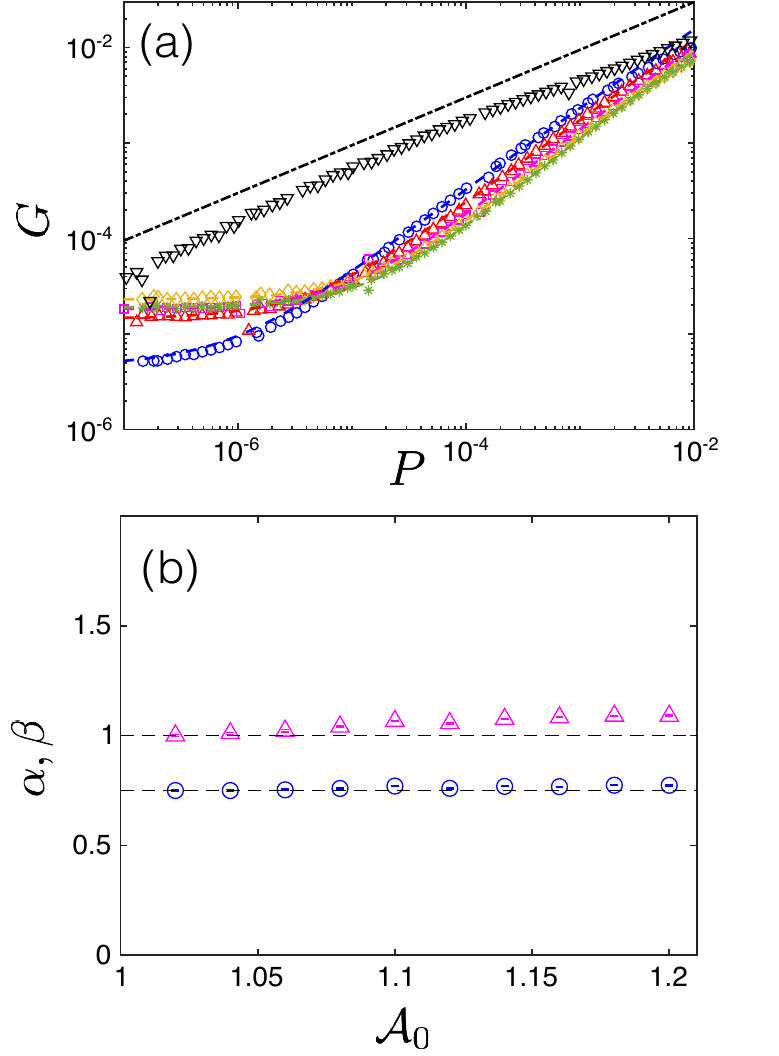}
    \caption{\textbf{Shear response differs in particle models with increasing deformability.} (a) Static shear modulus $G$ versus pressure $P$ for $N=256$ DP packings ($n_{\rm S} = 16$) with $\mathcal{A}_0 = 1.02$ (circles), $1.06$ (triangles), $1.1$ (squares), $1.14$ (diamonds), and $1.18$ (asterisks), and $N=256$ DPb packings ($n_{\rm S} = 16$) with $\mathcal{A}_0 = 1$ (downward triangles). The dashed lines are best fits to Eq.~\ref{eq:pg}. The dash-dotted line follows the scaling $G \sim P^{1/2}$. (b) Exponents $\alpha$ (triangles) and $\beta$ (circles) from Eq.~\ref{eq:pg} for the DP packings in (a) versus shape parameter $\mathcal{A}_0$. Horizontal lines indicate $\alpha = 1.0$ and $\beta =0.75$.}
    \label{fig:shearmodulus}
\end{figure}

To investigate the effect of particle deformability on bulk mechanical properties, we computed the static shear modulus $G$ for jammed packings of DP and DPb particles. Packings were compressed to a given pressure $P$, subjected to small, successive simple shear strain steps of size $\Delta \gamma$ with Lees-Edwards boundary conditions~\citep{sim:AllenOxford2017}, and the system was energy-minimized after each step. We measure $G = -d\Sigma_{xy}/d\gamma$, where $\Sigma_{xy}$ is the virial shear stress. We report $G$ averaged over an ensemble of at least $500$ configurations. In Fig.~\ref{fig:shearmodulus}, we show that, although DP packings contain collective low-frequency quartic modes, they possess $G > 0$ at low pressure~\citep{jamming:MailmanPRL2009,jamming:BoromandPRL2018}. In Appendix~\ref{sec:appendix:size}, we also show characteristic $N^{-1}$ scaling of $G$ in the $P \to 0$ limit~\citep{softp:GoodrichPRL2012}. We find in Fig.~\ref{fig:shearmodulus} (a) that $G(P)$ for DP packings over of wide range of $\mathcal{A}_0$ is well-approximated by the double-power-law functional form~\citep{jamming:VanderWerfPRL2020} used to describe the shear response of packings of soft frictionless spheres:
\begin{equation}\label{eq:pg}
    G = G_0 + \frac{a P^\alpha}{1 + cP^{\alpha - \beta}},
\end{equation}
where $a$ and $c$ are constants.
$G_0$ is the value in the $P\to 0$ limit, the exponent $\alpha$ controls the low $P$ response, and the exponent $\beta$ controls the high $P$ response. 

Values of $\alpha \approx 1$ and $\beta \approx 0.5$ have been reported in previous studies of jammed packings of frictionless spherical particles~\citep{jamming:OHernPRE2003,softp:GoodrichPRL2012}, frictional spherical particles~\citep{frictional:SomfaiPRE2007}, and bumpy particles~\citep{jamming:PapaPRL2013}. However, in Fig.~\ref{fig:shearmodulus} (b), we find that the large pressure scaling exponent $\beta \approx 0.75$ for DP packings. In Fig.~\ref{fig:shearmodulus} (a), we show that $\beta \approx 0.5$ for unbuckled DPb particles with $K_b = 10^{-2}$ and $\mathcal{A}_0 = 1.04$, although we do not observe a plateau at low pressures. This result indicates that the mechanical response for unbuckled DPb particles (with $\mathcal{A}_0 < \mathcal{A}_0^*$) is similar to that for rigid-shape spherical particles. 

Note also that $G(P)$ for packings of DPb particles possesses an even smaller scaling exponent at high pressures ($P \sim 10^{-2}$). At these pressures, particles are likely starting to deform from their energy-minimized states to fill in their surrounding Voronoi cell as the packing approaches confluence. However, to fully understand the root cause of these scaling exponents, future work is needed to connect the behavior of single packings to the ensemble average. Prior work on frictionless disks~\citep{jamming:VanderWerfPRL2020} showed that $G$ {\it decreases} with $P$ for individual packings with fixed contact networks. Only when the contact network changes does the shear modulus increase, leading to a scaling of $P^{1/2}$ when averaging over an ensemble of many configurations with many different contact changes. Understanding the power-law scaling of the ensemble-averaged $G(P)$ for deformable particles requires an analysis of how deformable particles break contacts in response to compression, as well as how the shear modulus varies with pressure when the interparticle contact network does not change~\citep{jamming:WangPRE2021}.

\section{Conclusions}~\label{sec:conclusions}
In this work, we have studied rigidity, the vibrational density of states, and the mechanical response in athermal, jammed solids composed of particles that can explicitly change shape to varying degrees. We can vary particle deformability by studying the Deformable Polygon (DP) model, where each particle has as many shape degrees of freedom as it has vertices, and the effectively rigid-shape DPb model, which includes bending energy. We also showed that DPb particles can buckle by increasing the preferred shape parameter $\mathcal{A}_0$ above a characteristic value ($\mathcal{A}_0^*$), which effectively provides DPb particles with an additional degree of freedom. When studying the rigidity of jammed packings of these particles, we find that DP and DPb particles typically do not jam at a standard isostatic point. Packings of DP particles have too few contacts for collective rigidity, but we find that there are higher-order terms in the potential energy expansion (i.e. quartic modes) that stabilize the packings. Packings of DPb particles below the buckling threshold jam at the expected isostatic point for rigid-shape bumpy particles, but buckled DPb particles jam with \emph{more} contacts than expected and seem to be \emph{hyper}static. If we assume that buckled DPb particles have an extra degree of freedom, however, these packings are hypostatic just as in packings of DP particles. Although we cannot reliably count missing contacts from the vibrational spectra for buckled DPb packings, we show that a heuristic counting criterion roughly validates the observation that buckled DPb particles have higher-order rigidity.

Analyzing the vibrational spectra in more detail, we show that the vibrational density of states $D(\omega)$ depends strongly on particle deformability. In particular, we show that the characteristic frequency of quartic modes for jammed DP particles scales \emph{inversely} with particle shape parameter, i.e. $\omega_0\sim\delta\mathcal{A}_0^{-1/3}$. This result differs from other systems with three vibrational bands, in particular jammed packings of ellipsoids~\citep{jamming:SchreckPRE2012} and breathing particles with size degrees of freedom~\citep{nonspherical:BritoPNAS2018}. We also find that collective shape degrees of freedom $S$ play an important role in the low-frequency vibrational response of DP particles across different shapes and with increasing compression. We show that packings of regular-polygon DPb particles in the rigid-shape-particle limit $K_c\to 0$ do not possess low-frequency collective shape degrees of freedom, whereas $S(\omega)>0$ for all $K_c$ for packings with buckled DPb particles. We further show that $G\sim P^{3/4}$ over a wide range of pressure for packings of DP particles for all shape parameters studied, which deviates from the power-law scaling for packings of rigid-shape spherical particles. In contrast, packings of regular-polygon DPb particles possess $G\sim P^{1/2}$ scaling. 

In all, our results show that explicit particle deformability qualitatively changes the linear response of soft solids. The bulk of these findings can be tested in experiments, either in non-contractile 2D monoloyers of epithelial cells (i.e. DP particles) or in soft, quasi-2D packings of hydrogel particles (i.e. regular-polygon DPb particles). The buckling phenomenon observed for DPb particles cannot easily be tested in an experiment, but we plan to carry out further theoretical studies of DPb buckling to gain a deeper understanding of quasi-deformability. Nevertheless, this work lays the foundation for understanding the vibrational and mechanical response in glassy systems of deformable particles, such as hopper flows of emulsion droplets~\citep{weeks:HongPRE2017,emulsions:GolovkovaSM2020} and motile tissues~\citep{collective:AjetiNatPhys2019}.

\section*{Acknowledgements}
We thank O. K. Damavindi, V. F. Hagh, C. D. Santangelo, and M. L. Manning for helpful discussions. We acknowledge support from NSF Grants No. BMMB-2029756 (J.T. and C.O.), No. CBET-2002782 (J.T. and C.O.), No. CBET-2002797 (M.S.), and No. CMMI-1463455 (M.S.) and NIH award No. 5U54CA210184-04 (D.W.). This work was also supported by the High Performance Computing facilities operated by Yale’s Center for Research Computing.

\appendix

\section{Particle buckling with bending energy}~\label{sec:appendix:buckling}
\begin{figure}
    \centering
    \includegraphics[width=0.5\textwidth]{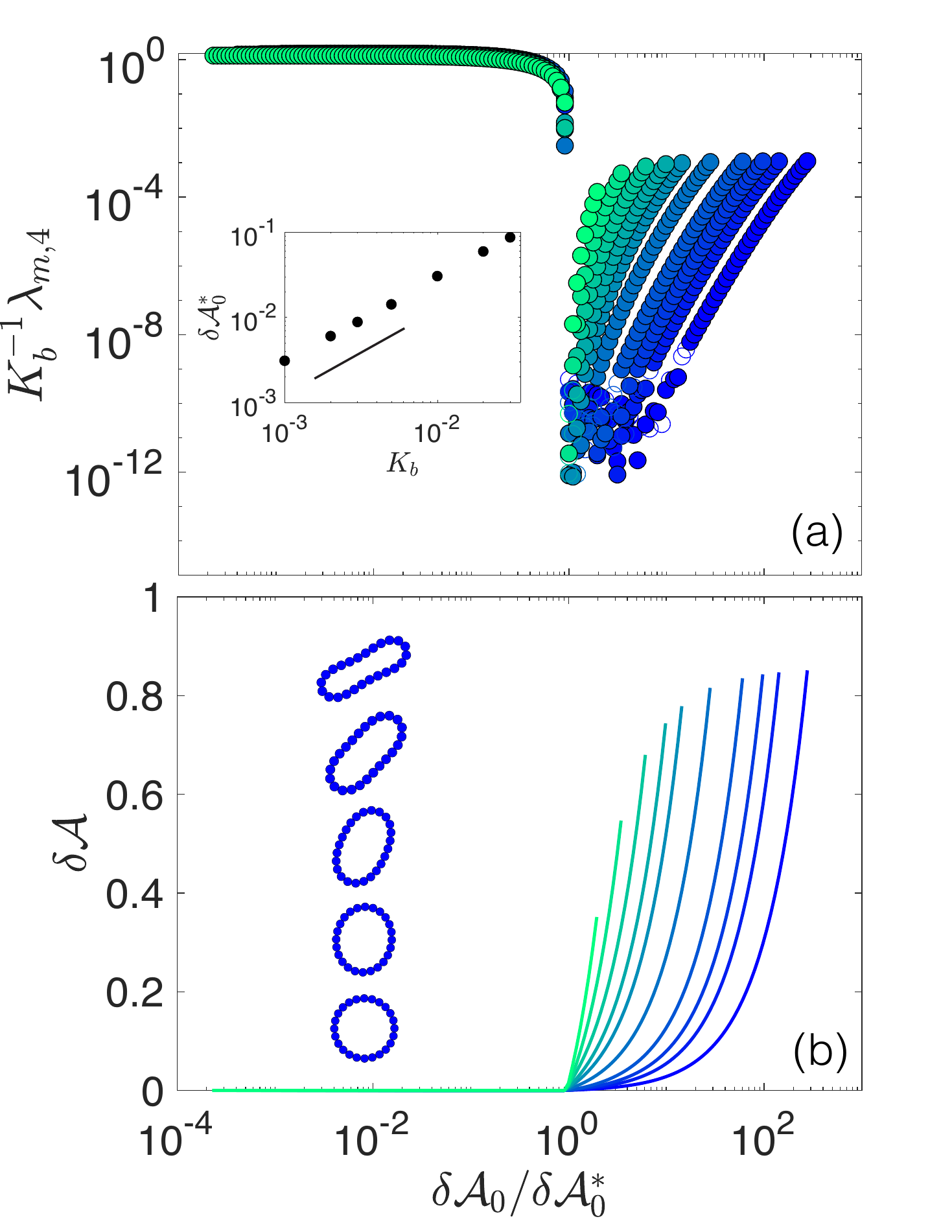}
    \caption{\textbf{Buckling in single DPb particles}. (a) The first non-trivial mode $\lambda_{m,4}$, scaled by the bending spring constant $K_b$, for DPb particles with $n = 24$ vertices as a function of the deviatoric preferred shape parameter $\delta \mathcal{A}_0 = \qty(\mathcal{A}_0 - \mathcal{A}_n)/\mathcal{A}_n$ and different $K_b$. Closed symbols indicate $\lambda_{m,4} > 0$, and open symbols indicate $\lambda_{m,4} < 0$. Blue colors represent smaller $K_b$, starting at $K_b = 10^{-3}$, and green colors represent larger $K_b$, ending with $K_b = 2\times 10^{-1}$. In both (a) and (b), the $x$-axis is scaled by $\delta \mathcal{A}_0^* = \qty(\mathcal{A}_0^* - \mathcal{A}_n)/\mathcal{A}_n$. The inset shows $\delta\mathcal{A}_0^*$ vs. $K_b$; $\mathcal{A}_0^*$ is defined as the preferred shape parameter when $\lambda_{m,4} < 10^{-8}$. (b) The \emph{true} deviatoric shape parameter $\mathcal{A} = p^2/4\pi a$ (i.e. $\delta \mathcal{A} = \qty(\mathcal{A} - \mathcal{A}_n)/\mathcal{A}_n$) of the buckled DPb particles as a function of $\delta A_0$. Colors are the same as in (a). The inset shows several representative particle shapes for $\delta\mathcal{A} = 0, 0.01, 0.1, 0.3,$ and $0.6$. These shapes are the same for any $K_b$ regardless of $\delta\mathcal{A}$.}
    \label{fig:singleDPbBuckling}
\end{figure}

In this Appendix, we demonstrate buckling of DPb particles by increasing the preferred shape parameter $\mathcal{A}_0$. In Fig.~\ref{fig:singleDPbBuckling} (a), we show that the first non-trivial mode of the single-particle vibrational spectrum of DPb particles decreases by several orders of magnitude at $\mathcal{A}_0^*$, which depends on $K_b$. In Fig.~\ref{fig:singleDPbBuckling} (b), we show that buckled particles transition from regular polygons with the true shape parameter $\mathcal{A} = \mathcal{A}_n$ to elongated, ellipsoidal particles with $\mathcal{A} > \mathcal{A}_n$ when $\mathcal{A}_0 > \mathcal{A}_0^*$. Note that occasionally, when sufficiently close to buckling, $\lambda_{m,4} < 0$, but $|\lambda_{m,4}|$ is close to numerical precision. These results suggest that, sufficiently close to buckling, DPb particles gain a degree of freedom and $\lambda_{m,4} \approx 0$. However, increasing ${\cal A}_0$ further causes the eigenvalue to grow in magnitude. While $\lambda_{m,4}$ remains significantly smaller than $K_b$, it is unclear whether DPb particles lose this degree of freedom at higher $\mathcal{A}_0$. 

\begin{figure}
    \centering
    \includegraphics[width=0.4\textwidth]{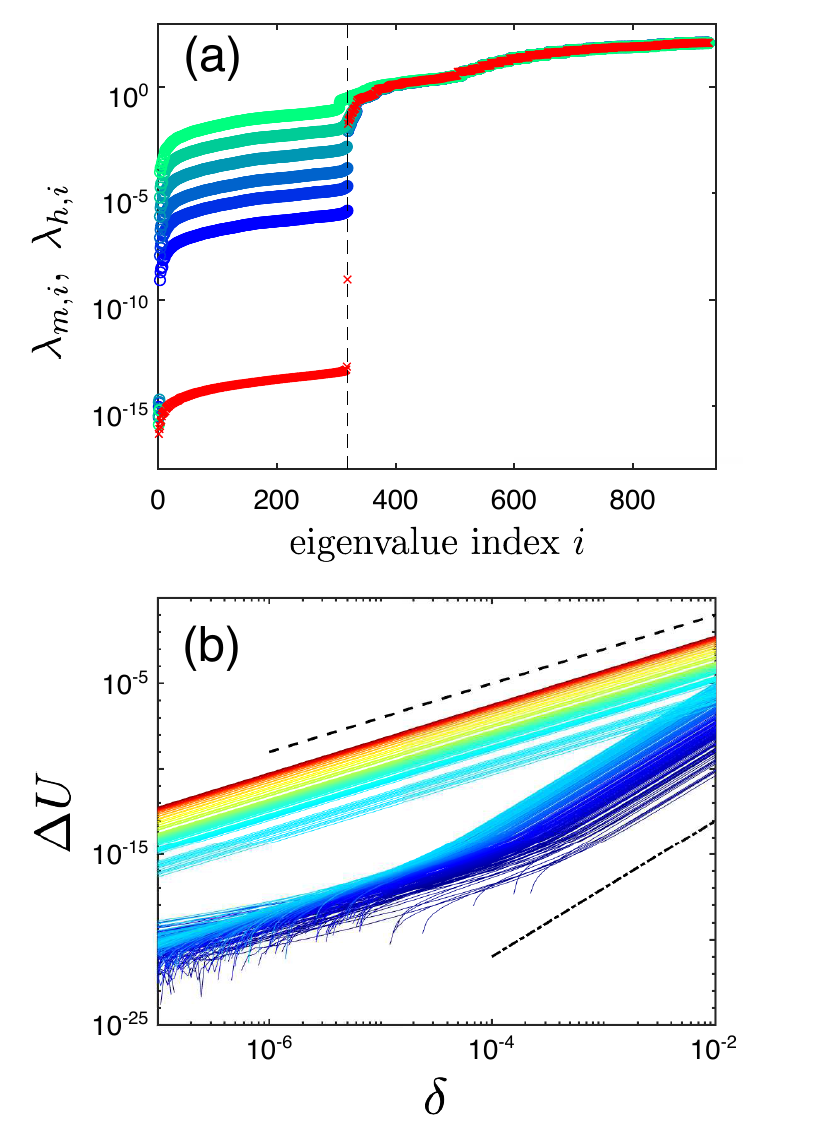}
    \caption{\textbf{Missing contacts in jammed packings of DP particles are stabilized by quartic modes.} (a) Eigenvalues of the dynamical ($\lambda_{m,i}$, circles) and stiffness ($\lambda_{h,i}$, crosses) matrices as a function of eigenvalue index $i$ for a configuration of $N = 16$ DP particles with $\mathcal{A}_0 = 1.02$. $\lambda_{h,i}$ do not depend on pressure, whereas $\lambda_{m,i}$ do. We show $\lambda_{m,i}$ for pressures from $P = 10^{-8}$ (blue) to $10^{-3}$ (green) separated by factors of $10$. The vertical line is placed at $2 + m$. (b) Change in potential energy $\Delta U = U - U_0$ for a system starting at an energy minimum $U_0$ for perturbations of size $\delta$ along the eigenmodes of the dynamical matrix for the systems in (a). Color indicates mode frequency, from blue (lowest) to red (highest). The dot-dashed line represents $\Delta U \sim \delta^4$, whereas the dashed line represents $\Delta U \sim \delta^2$. There is a one-to-one correspondence between the number of modes with quartic $\delta$-dependence in (b) and the pressure-dependent modes that are much larger than the stiffness matrix eigenvalues in (a).  }
    \label{fig:DpCounting}
\end{figure}

\begin{figure}
    \centering
    \includegraphics[width=0.5\textwidth]{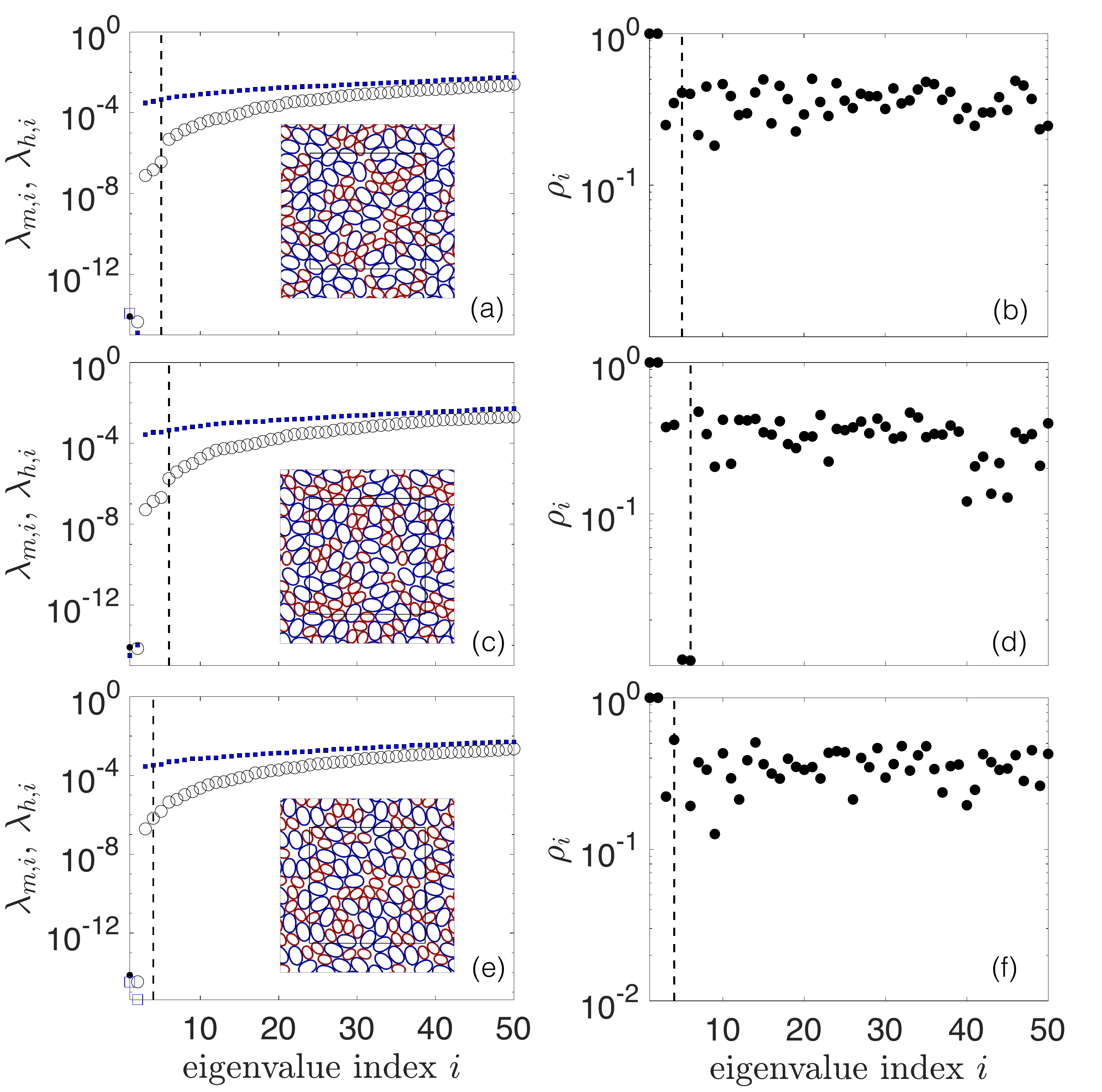}
    \caption{\textbf{Heuristic counting of missing constraints in packings of buckled DPb particles relies on gaps in the vibrational spectra and eigenmode participation ratios.} (a) The first $50$ eigenvalues of the dynamical matrix ($\lambda_m$, black circles) and stiffness matrix ($\lambda_h$, blue squares) for the jammed packings of $N = 64$ DPb particles ($\mathcal{A}_0 = 1.08 > \mathcal{A}_0^*$, $K_b = 10^{-2}$) shown in the inset. In the inset, small particles with $n_{\rm S} = 24$ vertices are drawn in red, and large particles with $n_{\rm L} = 34$ vertices are drawn in blue. Open blue squares (closed black circles) represent negative stiffness (dynamical) matrix eigenvalues. The vertical line drawn at index $i = 6$, which for this system was $2 + m$. Note that all systems considered here have no rattler particles, so there are only $2$ trivial zero modes. (b) The participation ratio $\rho_i$ from Eq.~\eqref{eq:participation} for each eigenmode of the packing in (a). The vertical line is also drawn at $i = 6$. (c) - (f) Same as (a) and (b), but for different jammed packings . Throughout, the vertical line is drawn at $2 + m$. }
    \label{fig:DpbCounting}
\end{figure}

\section{Higher-order constraints}~\label{sec:appendix:constraints}
\begin{figure}
    \centering
    \includegraphics[width=0.4\textwidth]{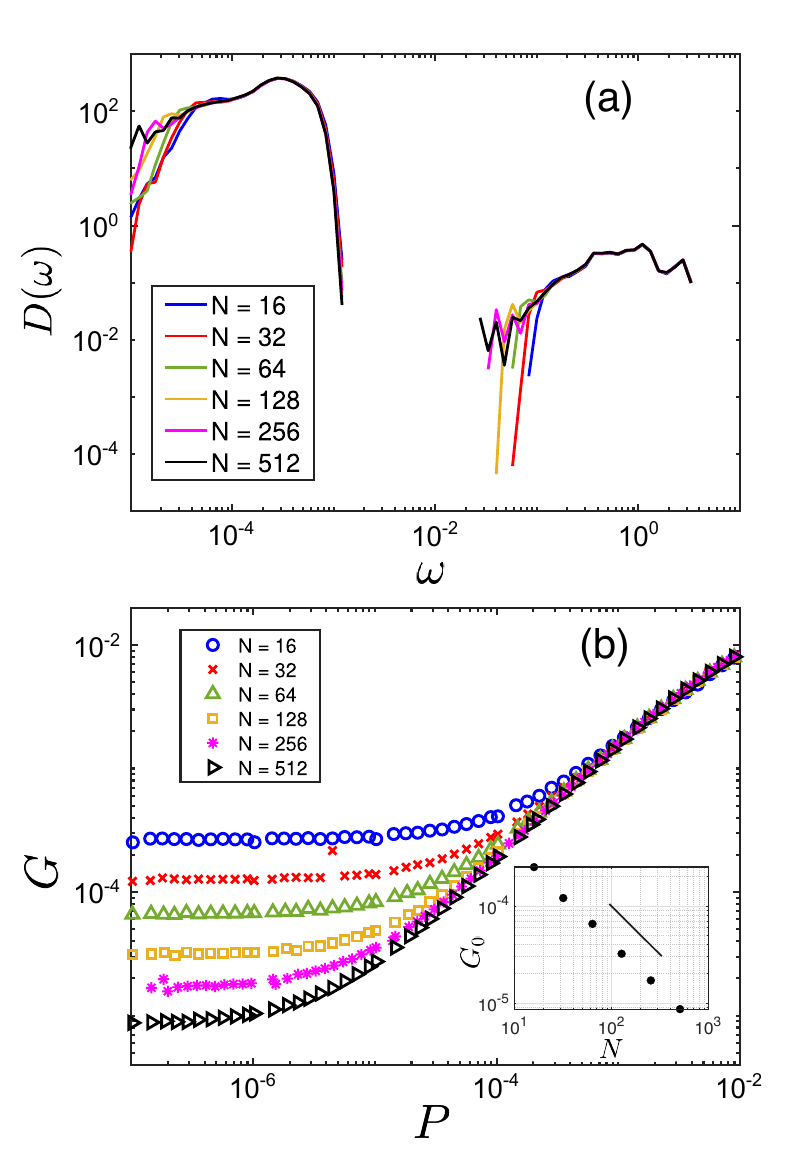}
    \caption{\textbf{System size dependence of the density of states and shear modulus}. (a) Vibrational density of states $D(\omega)$ versus eigenmode frequency $\omega$ for jammed packings of bidisperse DP particles with $\mathcal{A}_0 = 1.08$ from $N = 16$ to $N = 512$ with $n_{\rm S} = 16$. (b) Static shear modulus $G$ versus pressure $P$ for the same systems in (a). The inset shows $G_0$ (i.e. $G(P \to 0)$) versus system size $N$. The black solid line has the form $G_0 \sim N^{-1}$.  }
    \label{fig:sysSize}
\end{figure}
\begin{figure}
    \includegraphics[width=0.6\linewidth]{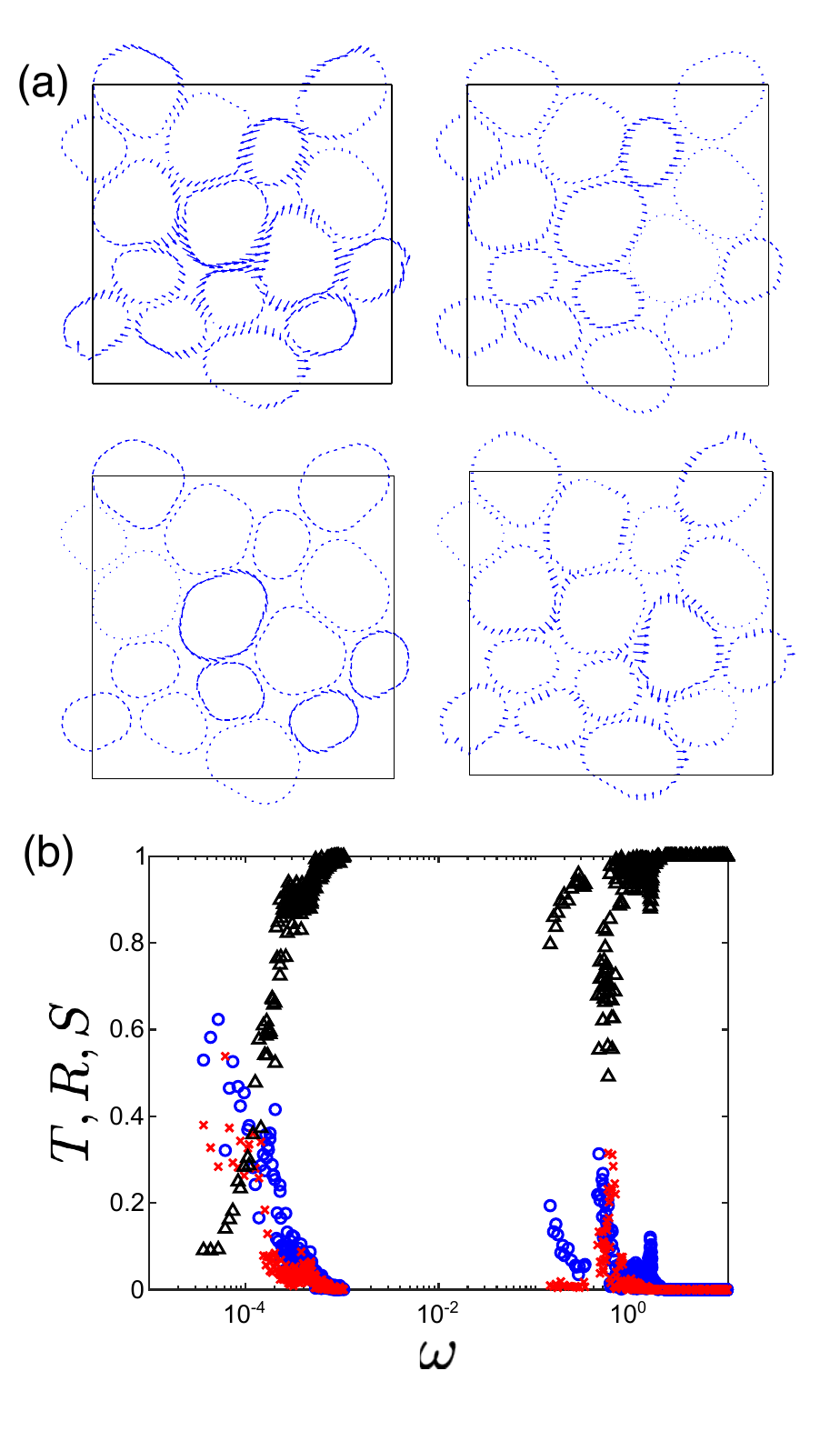}
    \caption{\textbf{Eigenvectors of the dynamical matrix can be decomposed into translational, rotational, and shape degrees of freedom.} (a) Example of an eigenmode of the dynamical matrix (upper left) with frequency $\omega \approx 1.2 \times 10^{-4}$ for a packing of $N = 16$ DP particles ($n_{\rm S} = 24$) with preferred shape parameter $\mathcal{A}_0 = 1.04$, and its decomposition into translation (upper right), rotation (lower left), and shape (lower right) degrees of freedom. The vectors are rescaled for clarity. (b) Contributions from translation $T$ (blue circles), rotation $R$ (red triangles), and shape $S$ (black squares) degrees of freedom for each eigenmode as a function of the eigenmode frequency $\omega$ for the DP particle packing in (a).}
    \label{fig:decomp}
\end{figure}

In this Appendix, we discuss how to count effective constraints in jammed packings of deformable particles by analyzing the vibrational eigenmodes. In general, the vibrational response of a packing of deformable particles is obtained by calculating the dynamical matrix $\mathcal{M}$ evaluated at a point of mechanical equilibrium, 
\begin{equation}
    \mathcal{M}_{ij\mu\nu} = \pdv{U}{\vec{r}_{j\nu}}{\vec{r}_{i\mu}},
\end{equation}
where $\vec{r}_{i\mu} = (x_{i\mu}, y_{i\mu})$ is the coordinate vector of vertex $i$ on particle $\mu$, and we bring the system to force balance ($\pdv*{U}{\vec{r}_{i\mu}} = \vec{0})$ before evaluating the matrix elements. Note that $\mathcal{M}_{ij\mu\nu}$ is a $d \times d$ block matrix. Consider the DP energy in Eq.~\ref{eq:dpEnergy} in the main text, where $U_a$, $U_l$, and $U_c$ represent the area, perimeter, and particle interaction contributions to the total potential energy, respectively. We can define dynamical matrices for each term, e.g. $\mathcal{M}_l$ is perimeter energy contribution to the dynamical matrix. Using the chain rule, first derivatives of $U_l$ with respect to the vertex coordinates can be written as
\begin{equation}
    \pdv{U_l}{\vec{r}_{i\mu}} = \sum_{\alpha = 1}^N\sum_{k=1}^{n_\alpha} \pdv{U_l}{l_{k\alpha}}\pdv{l_{k\alpha}}{\vec{r}_{i\mu}}
\end{equation}
since the perimeter energy depends on vertex coordinates through the edge length $l_{i\mu} = \abs{\vec{r}_{i+1,\mu} - \vec{r}_{i\mu}}$. Note that this applies to all terms in Eqs.~\ref{eq:dpEnergy} and~\ref{eq:ub} (e.g. $U_a$, $U_b$, and $U_c$) that depend on the degrees of freedom through geometric factors. The second derivatives have the following form:
\begin{equation}
    \begin{split}
        \pdv{U_l}{\vec{r}_{j\nu}}{\vec{r}_{i\mu}} &= \sum_{\alpha = 1}^N\sum_{k=1}^{n_\alpha} \pdv{\vec{r}_{j\nu}}\qty(\pdv{U_l}{l_{k\alpha}}\pdv{l_{k\alpha}}{\vec{r}_{i\mu}})\\ 
        &= \sum_{\alpha = 1}^N\sum_{k=1}^{n_\alpha}\qty( \pdv[2]{U_l}{l_{k\alpha}}\pdv{l_{k\alpha}}{\vec{r}_{j\nu}}\pdv{l_{k\alpha}}{\vec{r}_{i\mu}} + \pdv{U_l}{l_{k\alpha}}\pdv{l_{k\alpha}}{\vec{r}_{j\nu}}{\vec{r}_{i\mu}}).
    \end{split}
\end{equation}
The perimeter energy contribution to the dynamical matrix can be decomposed as $\mathcal{M}^l = \mathcal{H}^l - \mathcal{S}^l$, where
\begin{subequations}
    \begin{align}
        \mathcal{H}^l_{ij\mu\nu} &= \sum_{\alpha = 1}^N\sum_{k=1}^{n_\alpha}\pdv[2]{U_l}{l_{k\alpha}}\pdv{l_{k\alpha}}{\vec{r}_{j\nu}}\pdv{l_{k\alpha}}{\vec{r}_{i\mu}}\\
        \mathcal{S}^l_{ij\mu\nu} &= -\sum_{\alpha = 1}^N\sum_{k=1}^{n_\alpha} \pdv{U_l}{l_{k\alpha}}\pdv{l_{k\alpha}}{\vec{r}_{j\nu}}{\vec{r}_{i\mu}}
    \end{align}
\end{subequations}
are the ``stiffness" and ``stress" matrices, respectively~\citep{ellipse:DonevPRE2007,jamming:SchreckPRE2012}. The matrix $\mathcal{H}^l$ depends primarily on first derivatives of geometric factors (e.g. $l_{i\mu}$) with respect to vertex coordinates, while the matrix $\mathcal{S}^l$ depends primarily on first derivatives of the potential energy. Note that the stress and stiffness matrices of the entire potential energy can be computed by summing contributions from corresponding matrices defined by the different contributions to the potential energy, e.g. $\mathcal{H} = \mathcal{H}^a + \mathcal{H}^l + \mathcal{H}^c$ is the sum of the area, perimeter, and particle interaction energy contributions to $\mathcal{H}$.

Prior work has shown that hypostatic systems like jammed non-spherical particles~\citep{ellipse:DonevPRE2007,jamming:VanderWerfPRE2018,jamming:YuanSM2019} and underconstrained spring networks~\citep{rigidity:DamavindiArXiv2021} gain rigidity from higher-order terms in the potential energy that can stabilize multiple degrees of freedom. As noted in prior work~\citep{jamming:SchreckPRE2012}, these higher-order constraints can be seen as zero modes of the stiffness matrix $\mathcal{H}$, but non-zero modes of the dynamical matrix $\mathcal{M}$. In Fig.~\ref{fig:DpCounting} (a), we find that the number of missing contacts exactly equals the number of stiffness matrix eigenvalues $\lambda_h$ that are significantly smaller than dynamical matrix eigenvalues $\lambda_m$. Since $\mathcal{M} = \mathcal{H} - \mathcal{S}$, we expect that small stiffness contribution means that $\mathcal{M}$ is dominated by the stress matrix for these eigenvalues. Indeed, in Fig.~\ref{fig:DpCounting} (a) we also show that we can tune the magnitude of $\lambda_m$ by maintaining a fixed contact network but increasing the pressure. This pressure dependence of the low frequency eigenvalues of the dynamical matrix is also observed in packings of frictionless non-spherical particles~\citep{jamming:VanderWerfPRE2018}.

In hypostatic DP packings, we find that the extra degrees of freedom are constrained by so-called ``quartic modes". Consider an energy-minimized configuration such that the particle coordinates satisfy $\vec{R} = \vec{R}_0$. Perturbations of order $\delta$ are then written as $\vec{R} = \vec{R}_0 + \delta \vec{u}$, where $\vec{u}$ is the direction of the perturbation. The potential energy expanded about $\vec{R}_0$ to fourth order in the perturbation is
\begin{equation}
    \begin{split}
        U\qty(\vec{R}) &= U\qty(\vec{R}_0) + \delta \pdv{U}{R_i} u_i + \frac{\delta^2}{2}\pdv{U}{R_i}{R_j}u_i u_j\\
        &+ \frac{\delta^3}{6}\frac{\partial^3 U}{\partial R_i \partial R_j \partial R_k} u_i u_j u_k\\
        &+ \frac{\delta^4}{24}\frac{\partial^4 U}{\partial R_i \partial R_j \partial R_k \partial R_l} u_i u_j u_k u_l + \cdots,
    \end{split}
\end{equation}
where we sum over repeated indices, and all derivatives are evaluated at $\vec{R}_0$. The term linear in $\delta$ is $0$ when the system is at a potential energy minimum. If we choose $\vec{u}$ to be the $k$th orthonormal eigenvector of the dynamical matrix $\mathcal{M}$, the potential energy to second order in $\delta$ is $U = U_0 + \frac{1}{2}\lambda_{m,k}\delta^2$. However, in Fig.~\ref{fig:DpCounting} (b), we find that the potential energy scales as $U \sim \delta^4$, which is consistent with the observation that $\lambda_{m,k}$ are small, but the quartic terms are non-negligible. Similar behavior is observed in jammed packings of non-spherical particles~\citep{jamming:MailmanPRL2009,jamming:SchreckPRE2012,jamming:VanderWerfPRE2018}. 

While we can easily identify the number of higher-order contacts from the dynamical matrix eigenspectra for packings of DP particles, the same is not true for packings of DPb particles. In Fig.~\ref{fig:DpbCounting}, we show that the stiffness matrix eigenvalues $\lambda_h$ are \emph{larger} than the dynamical matrix eigenvalues $\lambda_m$ for packings of buckled DPb particles.  Given that buckled DPb particles have an extra degree of freedom, we investigate whether there is a signature in the eigenmodes with indexes below $m = 4N' - 1 - N_{\rm vv}$, which corresponds to the number of missing contacts in packings of $N'$ non-rattler particles.

We develop the following heuristic for packings of DPb particles: first we check if there is a gap of at least a factor of $10$ between the $m$th non-trivial eigenmode of the dynamical matrix and the next mode, since there is a gap between quartic and quadratic modes for packings of DP particles. If a gap in the eigenspectra of the dynamical matrix is not present, we calculate the participation ratio of each normal mode $k$, 
\begin{equation}\label{eq:participation}
    \rho_k = \frac{\qty(\sum_{\mu=1}^N\sum_{i=1}^{n_\mu} \abs{\vec{V}_{i\mu,k}}^2)^2 }{N\sum_{\mu=1}^N \sum_{i=1}^{n_\mu} \abs{\vec{V}_{i\mu,k}}^4},
\end{equation}
where $\vec{V}_{i\mu, k}$ is the displacement direction of the $i$th vertex on the $\mu$th particle in mode $k$. If we observe a gap in the participation ratio of at least a factor of $10$ between the $m$th non-trivial mode and the next mode, we assume that we have identified the $m$ higher-order modes correctly. We use the participation ratio gap because, in general, the participation ratio decreases with increasing eigenmode frequency. However, for DP particles, the highest-frequency quartic mode is usually localized whereas the lowest-frequency quadratic mode is delocalized. If neither of these two conditions is satisfied, we assume that there are no higher-order modes and no missing contacts. We emphasize that these thresholds are {\it ad hoc}, as the root cause of higher-order stability in jammed packings of buckled DPb particles is still an active area of research.

In Fig.~\ref{fig:DpbCounting}, we show the outcome of this heuristic counting for several example configurations of $N = 64$ buckled DPb particles at jamming onset. For example, in Fig.~\ref{fig:DpbCounting} (a) and (b), we find that a gap at the $m$th non-trivial mode appears, allowing us to identify these modes as higher-order constraints in analogy with packings of DP particles. We also can correctly identify higher-order constraints using the participation ratio as shown in Fig.~\ref{fig:DpbCounting} (c) and (d). However, we show in Fig.~\ref{fig:isostaticity} (b) and in Fig.~\ref{fig:DpbCounting} (e) and (f) that there are several cases with missing contacts $m > 0$, but we cannot identify the missing contacts by analyzing the eigenvalue spectra or participation ratios. The fact that missing contacts cannot be counted consistently underscores the difficulty in identifying higher-order constraints in these jammed systems.

\section{System size dependence of the vibrational density of states and shear modulus}~\label{sec:appendix:size}

In this Appendix, we investigate the system size dependence in the vibrational density of states $D(\omega)$ and static shear modulus $G$ of jammed packings of DP particles. In Fig.~\ref{fig:sysSize} (a), we show $D(\omega)$ for multiple system sizes spanning $N = 16$ to $N = 512$ with $n_{\rm S} = 16$ for the preferred shape parameter $\mathcal{A}_0 = 1.08$. We see little change in the structure of $D(\omega)$ except for more low-frequency quartic modes for larger system sizes, though this does not seem to change the peaks in $D(\omega)$. We do however see system-size dependence in the static shear modulus as shown in Fig.~\ref{fig:sysSize} (b) in the low-pressure limit. At high pressures, $G$ collapses across all system sizes studied, but as $P \to 0$ we show in the Fig.~\ref{fig:sysSize} (b) inset that $G(P \to 0) \sim N^{-1}$, which has been observed in previous work on deformable particles~\citep{jamming:BoromandPRL2018}. 

\section{Mode decomposition}~\label{sec:appendix:decomp}
In this Appendix, we will show in detail how to decompose the eigenmodes of the dynamical matrix into contributions from particle translation, rotation, and shape degrees of freedom. We consider a packing of $N$ deformable particles, where each particle $\mu$ has a center of mass located at $\vec{c}_\mu = n_\mu^{-1}\sum_{i=1}^{n_\mu} \vec{r}_{i\mu}$. Let $\vec{V}^j$ be the unit vector corresponding to the $j$th eigenmode of the dynamical matrix $\mathcal{M}$ in Cartesian coordinates. Individual components of $\vec{V}^j$ are arranged such that the components from $2n_{\mu-1}$ to $2n_\mu$ are the $n_{\mu}$ $x$-coordinates followed by the $n_{\mu}$ $y$-coordinates for the $\mu$th deformable particle. We can write three unit vectors to describe translation ($\vu{u}_{\mu,x}$, $\vu{u}_{\mu,y})$ and rotation ($\vu{u}_{\mu,r}$) about the center of mass of the $\mu$th particle as follows: 
\begin{widetext}
\begin{equation}
    \vu{u}_{\mu, x} = \frac{\vec{u}_{\mu, x}}{|\vec{u}_{\mu, x}|}, \vec{u}_{\mu, x} = (\underbrace{0, \ldots, 0}_{\text{1 to ($\mu - 1$)}}, \underbrace{1, \ldots, 1}_{\text{$\mu$th particle $x$}}, \underbrace{0, \ldots, 0}_{\text{$\mu$th particle $y$}}, \underbrace{0, \ldots, 0}_{\text{($\mu$+1) to $N$}})
\end{equation}
\begin{equation}
    \vu{u}_{\mu, y} = \frac{\vec{u}_{\mu, y}}{|\vec{u}_{\mu, y}|}, \vec{u}_{\mu, y} = (\underbrace{0, \ldots, 0}_{\text{$1$ to ($\mu - 1$) }}, \underbrace{0, \ldots, 0}_{\text{$\mu$th particle $x$}}, \underbrace{1, \ldots, 1}_{\text{$\mu$th particle $y$}}, \underbrace{0, \ldots, 0}_{\text{$(\mu+1)$ to $N$}})
\end{equation}
\begin{equation}
\begin{split}
    \vu{u}_{\mu, r} = \frac{\vec{u}_{\mu, r}}{|\vec{u}_{\mu, r}|}, \vec{u}_{\mu, r} = (\underbrace{0, \ldots, 0}_{\text{$1$ to ($\mu - 1$) }}, \underbrace{-(y_{1\mu} - c_{\mu, y}), \ldots, -(y_{n_{\mu}\mu} - c_{\mu, y})}_{\text{$\mu$th particle $x$}}, \\ \underbrace{x_{1\mu} - c_{\mu, x}, \ldots, x_{n_{\mu}\mu} - c_{\mu, x}}_{\text{$\mu$th particle $y$}}, \underbrace{0, \ldots, 0}_{\text{$(\mu+1)$ to $N$}}).
\end{split}
\end{equation}
\end{widetext}
By defining the coefficients,
\begin{align}
    p_{\mu, x}^j &= \vec{V}^j \cdot \vu{u}_{\mu, x}\\
    p_{\mu, y}^j &= \vec{V}^j \cdot \vu{u}_{\mu, y}\\
    p_{\mu, r}^j &= \vec{V}^j \cdot \vu{u}_{\mu, r},
\end{align}
we can rewrite the eigenvector $\vec{V}^j$ as
\begin{equation}
    \vec{V}^j = \sum_{\mu = 1}^{N} p_{\mu, x}^j \vu{u}_{\mu, x} + \sum_{\mu = 1}^{N} p_{\mu, y}^j \vu{u}_{\mu, y} + \sum_{\mu = 1}^{N} p_{\mu, r}^j \vu{u}_{\mu, r} + \vec{V}_s^j,
\end{equation}
where $\vec{V}_s^j$ is the vector that remains after subtracting the particle translations and rotation out of $\vec{V}_j$. By applying this decomposition, we can express each eigenmode as the sum of particle translations, rotation, and shape deformations.  We show an example of an eigenmode decomposition in Fig.~\ref{fig:decomp} (a) for a packing of DP particles.

With these coefficients, we can define the fraction of the translational ($T^j$) and rotational ($R^j$) content in the $j$th eigenmode of the dynamical matrix as:
\begin{align}
    T^j &= \sum_{\mu = 1}^{N} \qty[\qty(p_{\mu, x}^j)^2 + \qty(p_{\mu, y}^j)^2]\\
    R^j &= \sum_{\mu = 1}^{N} \qty(p_{\mu, r}^j)^2.
\end{align}
Since we obtain $p_{\mu, x}^j, p_{\mu, y}^j$, and $p_{\mu, r}^j$ from unit vectors, $S^j = 1 - T^j - R^j$ gives the contribution of the shape degrees of freedom to the $j$th eigenmode. We show $T^j, R^j$, and $S^j$ for a jammed packing of $N = 16$ DP particles with preferred shape parameter $\mathcal{A}_0 = 1.04$ in Fig.~\ref{fig:decomp} (b) as a function of frequency $\omega$, as well as just the $S$ modes in Figs.~\ref{fig:modeProjection} and~\ref{fig:rigidLimit}. We find that for the quartic modes ($\omega < 10^{-2}$), the shape contribution $S$ increases with $\omega$, while $T$ and $R$ decrease. For the quadratic modes ($\omega > 10^{-2}$), the contribution from $S$ is large, since higher frequency modes tend to deform the particle shape rather than give rise to translation or rotation.

\bibliography{main}

\begin{thebibliography}{54}%
\makeatletter
\providecommand \@ifxundefined [1]{%
 \@ifx{#1\undefined}
}%
\providecommand \@ifnum [1]{%
 \ifnum #1\expandafter \@firstoftwo
 \else \expandafter \@secondoftwo
 \fi
}%
\providecommand \@ifx [1]{%
 \ifx #1\expandafter \@firstoftwo
 \else \expandafter \@secondoftwo
 \fi
}%
\providecommand \natexlab [1]{#1}%
\providecommand \enquote  [1]{``#1''}%
\providecommand \bibnamefont  [1]{#1}%
\providecommand \bibfnamefont [1]{#1}%
\providecommand \citenamefont [1]{#1}%
\providecommand \href@noop [0]{\@secondoftwo}%
\providecommand \href [0]{\begingroup \@sanitize@url \@href}%
\providecommand \@href[1]{\@@startlink{#1}\@@href}%
\providecommand \@@href[1]{\endgroup#1\@@endlink}%
\providecommand \@sanitize@url [0]{\catcode `\\12\catcode `\$12\catcode
  `\&12\catcode `\#12\catcode `\^12\catcode `\_12\catcode `\%12\relax}%
\providecommand \@@startlink[1]{}%
\providecommand \@@endlink[0]{}%
\providecommand \url  [0]{\begingroup\@sanitize@url \@url }%
\providecommand \@url [1]{\endgroup\@href {#1}{\urlprefix }}%
\providecommand \urlprefix  [0]{URL }%
\providecommand \Eprint [0]{\href }%
\providecommand \doibase [0]{http://dx.doi.org/}%
\providecommand \selectlanguage [0]{\@gobble}%
\providecommand \bibinfo  [0]{\@secondoftwo}%
\providecommand \bibfield  [0]{\@secondoftwo}%
\providecommand \translation [1]{[#1]}%
\providecommand \BibitemOpen [0]{}%
\providecommand \bibitemStop [0]{}%
\providecommand \bibitemNoStop [0]{.\EOS\space}%
\providecommand \EOS [0]{\spacefactor3000\relax}%
\providecommand \BibitemShut  [1]{\csname bibitem#1\endcsname}%
\let\auto@bib@innerbib\@empty
%</preamble>
\bibitem [{\citenamefont {Durian}(1995)}]{flow:DurianPRL1995}%
  \BibitemOpen
  \bibfield  {author} {\bibinfo {author} {\bibfnamefont {D.~J.}\ \bibnamefont
  {Durian}},\ }\href {\doibase 10.1103/PhysRevLett.75.4780} {\bibfield
  {journal} {\bibinfo  {journal} {Phys. Rev. Lett.}\ }\textbf {\bibinfo
  {volume} {75}},\ \bibinfo {pages} {4780} (\bibinfo {year}
  {1995})}\BibitemShut {NoStop}%
\bibitem [{\citenamefont {van Hecke}(2009)}]{softp:vanHeckeJPCM2009}%
  \BibitemOpen
  \bibfield  {author} {\bibinfo {author} {\bibfnamefont {M.}~\bibnamefont {van
  Hecke}},\ }\href {\doibase 10.1088/0953-8984/22/3/033101} {\bibfield
  {journal} {\bibinfo  {journal} {J. Phys.: Condens. Matter}\ }\textbf
  {\bibinfo {volume} {22}},\ \bibinfo {pages} {033101} (\bibinfo {year}
  {2009})}\BibitemShut {NoStop}%
\bibitem [{\citenamefont {Bolton}\ and\ \citenamefont
  {Weaire}(1990)}]{foams:BoltonPRL1990}%
  \BibitemOpen
  \bibfield  {author} {\bibinfo {author} {\bibfnamefont {F.}~\bibnamefont
  {Bolton}}\ and\ \bibinfo {author} {\bibfnamefont {D.}~\bibnamefont
  {Weaire}},\ }\href {\doibase 10.1103/PhysRevLett.65.3449} {\bibfield
  {journal} {\bibinfo  {journal} {Phys. Rev. Lett.}\ }\textbf {\bibinfo
  {volume} {65}},\ \bibinfo {pages} {3449} (\bibinfo {year}
  {1990})}\BibitemShut {NoStop}%
\bibitem [{\citenamefont {Bertho}\ \emph {et~al.}(2006)\citenamefont {Bertho},
  \citenamefont {Becco},\ and\ \citenamefont
  {Vandewalle}}]{hopper:BerthoPRE2006}%
  \BibitemOpen
  \bibfield  {author} {\bibinfo {author} {\bibfnamefont {Y.}~\bibnamefont
  {Bertho}}, \bibinfo {author} {\bibfnamefont {C.}~\bibnamefont {Becco}}, \
  and\ \bibinfo {author} {\bibfnamefont {N.}~\bibnamefont {Vandewalle}},\
  }\href {\doibase 10.1103/PhysRevE.73.056309} {\bibfield  {journal} {\bibinfo
  {journal} {Phys. Rev. E}\ }\textbf {\bibinfo {volume} {73}},\ \bibinfo
  {pages} {056309} (\bibinfo {year} {2006})}\BibitemShut {NoStop}%
\bibitem [{\citenamefont
  {Princen}(1983)}]{emulsion:PrincenJColloidInterSci1983}%
  \BibitemOpen
  \bibfield  {author} {\bibinfo {author} {\bibfnamefont {H.}~\bibnamefont
  {Princen}},\ }\href {\doibase https://doi.org/10.1016/0021-9797(83)90323-5}
  {\bibfield  {journal} {\bibinfo  {journal} {J. Colloid. Interf Sci.}\
  }\textbf {\bibinfo {volume} {91}},\ \bibinfo {pages} {160 } (\bibinfo {year}
  {1983})}\BibitemShut {NoStop}%
\bibitem [{\citenamefont {Boromand}\ \emph {et~al.}(2019)\citenamefont
  {Boromand}, \citenamefont {Signoriello}, \citenamefont {Lowensohn},
  \citenamefont {Orellana}, \citenamefont {Weeks}, \citenamefont {Ye},
  \citenamefont {Shattuck},\ and\ \citenamefont
  {O{'}Hern}}]{jamming:BoromandSM2019}%
  \BibitemOpen
  \bibfield  {author} {\bibinfo {author} {\bibfnamefont {A.}~\bibnamefont
  {Boromand}}, \bibinfo {author} {\bibfnamefont {A.}~\bibnamefont
  {Signoriello}}, \bibinfo {author} {\bibfnamefont {J.}~\bibnamefont
  {Lowensohn}}, \bibinfo {author} {\bibfnamefont {C.~S.}\ \bibnamefont
  {Orellana}}, \bibinfo {author} {\bibfnamefont {E.~R.}\ \bibnamefont {Weeks}},
  \bibinfo {author} {\bibfnamefont {F.}~\bibnamefont {Ye}}, \bibinfo {author}
  {\bibfnamefont {M.~D.}\ \bibnamefont {Shattuck}}, \ and\ \bibinfo {author}
  {\bibfnamefont {C.~S.}\ \bibnamefont {O{'}Hern}},\ }\href {\doibase
  10.1039/C9SM00775J} {\bibfield  {journal} {\bibinfo  {journal} {Soft Matter}\
  }\textbf {\bibinfo {volume} {15}},\ \bibinfo {pages} {5854} (\bibinfo {year}
  {2019})}\BibitemShut {NoStop}%
\bibitem [{\citenamefont {Smith}\ \emph {et~al.}(2017)\citenamefont {Smith},
  \citenamefont {Davit}, \citenamefont {Osborne}, \citenamefont {Kim},
  \citenamefont {Foster},\ and\ \citenamefont
  {Pitt-Francis}}]{biofilm:SmithPNAS2017}%
  \BibitemOpen
  \bibfield  {author} {\bibinfo {author} {\bibfnamefont {W.~P.~J.}\
  \bibnamefont {Smith}}, \bibinfo {author} {\bibfnamefont {Y.}~\bibnamefont
  {Davit}}, \bibinfo {author} {\bibfnamefont {J.~M.}\ \bibnamefont {Osborne}},
  \bibinfo {author} {\bibfnamefont {W.}~\bibnamefont {Kim}}, \bibinfo {author}
  {\bibfnamefont {K.~R.}\ \bibnamefont {Foster}}, \ and\ \bibinfo {author}
  {\bibfnamefont {J.~M.}\ \bibnamefont {Pitt-Francis}},\ }\href {\doibase
  10.1073/pnas.1613007114} {\bibfield  {journal} {\bibinfo  {journal} {Proc.
  Natl. Acad. Sci. USA}\ }\textbf {\bibinfo {volume} {114}},\ \bibinfo {pages}
  {E280} (\bibinfo {year} {2017})}\BibitemShut {NoStop}%
\bibitem [{\citenamefont {Beroz}\ \emph {et~al.}(2018)\citenamefont {Beroz},
  \citenamefont {Yan}, \citenamefont {Meir}, \citenamefont {Sabass},
  \citenamefont {Stone}, \citenamefont {Bassler},\ and\ \citenamefont
  {Wingreen}}]{biofilm:BerozNatPhys2018}%
  \BibitemOpen
  \bibfield  {author} {\bibinfo {author} {\bibfnamefont {F.}~\bibnamefont
  {Beroz}}, \bibinfo {author} {\bibfnamefont {J.}~\bibnamefont {Yan}}, \bibinfo
  {author} {\bibfnamefont {Y.}~\bibnamefont {Meir}}, \bibinfo {author}
  {\bibfnamefont {B.}~\bibnamefont {Sabass}}, \bibinfo {author} {\bibfnamefont
  {H.~A.}\ \bibnamefont {Stone}}, \bibinfo {author} {\bibfnamefont {B.~L.}\
  \bibnamefont {Bassler}}, \ and\ \bibinfo {author} {\bibfnamefont {N.~S.}\
  \bibnamefont {Wingreen}},\ }\href {\doibase 10.1038/s41567-018-0170-4}
  {\bibfield  {journal} {\bibinfo  {journal} {Nat. Phys.}\ }\textbf {\bibinfo
  {volume} {14}},\ \bibinfo {pages} {954} (\bibinfo {year} {2018})}\BibitemShut
  {NoStop}%
\bibitem [{\citenamefont {Park}\ \emph {et~al.}(2015)\citenamefont {Park},
  \citenamefont {Kim}, \citenamefont {Bi}, \citenamefont {Mitchel},
  \citenamefont {Qazvini}, \citenamefont {Tantisira}, \citenamefont {Park},
  \citenamefont {McGill}, \citenamefont {Kim}, \citenamefont {Gweon},
  \citenamefont {Notbohm}, \citenamefont {Steward~Jr}, \citenamefont {Burger},
  \citenamefont {Randell}, \citenamefont {Kho}, \citenamefont {Tambe},
  \citenamefont {Hardin}, \citenamefont {Shore}, \citenamefont {Israel},
  \citenamefont {Weitz}, \citenamefont {Tschumperlin}, \citenamefont {Henske},
  \citenamefont {Weiss}, \citenamefont {Manning}, \citenamefont {Butler},
  \citenamefont {Drazen},\ and\ \citenamefont
  {Fredberg}}]{collective:ParkNatMaterials2015}%
  \BibitemOpen
  \bibfield  {author} {\bibinfo {author} {\bibfnamefont {J.-A.}\ \bibnamefont
  {Park}}, \bibinfo {author} {\bibfnamefont {J.~H.}\ \bibnamefont {Kim}},
  \bibinfo {author} {\bibfnamefont {D.}~\bibnamefont {Bi}}, \bibinfo {author}
  {\bibfnamefont {J.~A.}\ \bibnamefont {Mitchel}}, \bibinfo {author}
  {\bibfnamefont {N.~T.}\ \bibnamefont {Qazvini}}, \bibinfo {author}
  {\bibfnamefont {K.}~\bibnamefont {Tantisira}}, \bibinfo {author}
  {\bibfnamefont {C.~Y.}\ \bibnamefont {Park}}, \bibinfo {author}
  {\bibfnamefont {M.}~\bibnamefont {McGill}}, \bibinfo {author} {\bibfnamefont
  {S.-H.}\ \bibnamefont {Kim}}, \bibinfo {author} {\bibfnamefont
  {B.}~\bibnamefont {Gweon}}, \bibinfo {author} {\bibfnamefont
  {J.}~\bibnamefont {Notbohm}}, \bibinfo {author} {\bibfnamefont
  {R.}~\bibnamefont {Steward~Jr}}, \bibinfo {author} {\bibfnamefont
  {S.}~\bibnamefont {Burger}}, \bibinfo {author} {\bibfnamefont {S.~H.}\
  \bibnamefont {Randell}}, \bibinfo {author} {\bibfnamefont {A.~T.}\
  \bibnamefont {Kho}}, \bibinfo {author} {\bibfnamefont {D.~T.}\ \bibnamefont
  {Tambe}}, \bibinfo {author} {\bibfnamefont {C.}~\bibnamefont {Hardin}},
  \bibinfo {author} {\bibfnamefont {S.~A.}\ \bibnamefont {Shore}}, \bibinfo
  {author} {\bibfnamefont {E.}~\bibnamefont {Israel}}, \bibinfo {author}
  {\bibfnamefont {D.~A.}\ \bibnamefont {Weitz}}, \bibinfo {author}
  {\bibfnamefont {D.~J.}\ \bibnamefont {Tschumperlin}}, \bibinfo {author}
  {\bibfnamefont {E.~P.}\ \bibnamefont {Henske}}, \bibinfo {author}
  {\bibfnamefont {S.~T.}\ \bibnamefont {Weiss}}, \bibinfo {author}
  {\bibfnamefont {M.~L.}\ \bibnamefont {Manning}}, \bibinfo {author}
  {\bibfnamefont {J.~P.}\ \bibnamefont {Butler}}, \bibinfo {author}
  {\bibfnamefont {J.~M.}\ \bibnamefont {Drazen}}, \ and\ \bibinfo {author}
  {\bibfnamefont {J.~J.}\ \bibnamefont {Fredberg}},\ }\href {\doibase
  10.1038/nmat4357} {\bibfield  {journal} {\bibinfo  {journal} {Nat. Mater.}\
  }\textbf {\bibinfo {volume} {14}},\ \bibinfo {pages} {1040} (\bibinfo {year}
  {2015})}\BibitemShut {NoStop}%
\bibitem [{\citenamefont {Trepat}\ and\ \citenamefont
  {Sahai}(2018)}]{collective:TrepatNatPhys2018}%
  \BibitemOpen
  \bibfield  {author} {\bibinfo {author} {\bibfnamefont {X.}~\bibnamefont
  {Trepat}}\ and\ \bibinfo {author} {\bibfnamefont {E.}~\bibnamefont {Sahai}},\
  }\href {\doibase 10.1038/s41567-018-0194-9} {\bibfield  {journal} {\bibinfo
  {journal} {Nat. Phys.}\ }\textbf {\bibinfo {volume} {14}},\ \bibinfo {pages}
  {671} (\bibinfo {year} {2018})}\BibitemShut {NoStop}%
\bibitem [{\citenamefont {Lecuit}\ and\ \citenamefont
  {Lenne}(2007)}]{shape:LecuitNatRevMolCellBio2007}%
  \BibitemOpen
  \bibfield  {author} {\bibinfo {author} {\bibfnamefont {T.}~\bibnamefont
  {Lecuit}}\ and\ \bibinfo {author} {\bibfnamefont {P.-F.}\ \bibnamefont
  {Lenne}},\ }\href {\doibase 10.1038/nrm2222} {\bibfield  {journal} {\bibinfo
  {journal} {Nat. Rev. Mol. Cell Bio.}\ }\textbf {\bibinfo {volume} {8}},\
  \bibinfo {pages} {633} (\bibinfo {year} {2007})}\BibitemShut {NoStop}%
\bibitem [{\citenamefont {Murrell}\ \emph {et~al.}(2015)\citenamefont
  {Murrell}, \citenamefont {Oakes}, \citenamefont {Lenz},\ and\ \citenamefont
  {Gardel}}]{actomyo:MurrellNatRevMolCellBio2015}%
  \BibitemOpen
  \bibfield  {author} {\bibinfo {author} {\bibfnamefont {M.}~\bibnamefont
  {Murrell}}, \bibinfo {author} {\bibfnamefont {P.~W.}\ \bibnamefont {Oakes}},
  \bibinfo {author} {\bibfnamefont {M.}~\bibnamefont {Lenz}}, \ and\ \bibinfo
  {author} {\bibfnamefont {M.~L.}\ \bibnamefont {Gardel}},\ }\href {\doibase
  10.1038/nrm4012} {\bibfield  {journal} {\bibinfo  {journal} {Nat. Rev. Mol.
  Cell Bio.}\ }\textbf {\bibinfo {volume} {16}},\ \bibinfo {pages} {486}
  (\bibinfo {year} {2015})}\BibitemShut {NoStop}%
\bibitem [{\citenamefont {Jolly}\ \emph {et~al.}(2015)\citenamefont {Jolly},
  \citenamefont {Boareto}, \citenamefont {Huang}, \citenamefont {Jia},
  \citenamefont {Lu}, \citenamefont {Ben-Jacob}, \citenamefont {Onuchic},\ and\
  \citenamefont {Levine}}]{cancer:JollyFrontOncol2015}%
  \BibitemOpen
  \bibfield  {author} {\bibinfo {author} {\bibfnamefont {M.~K.}\ \bibnamefont
  {Jolly}}, \bibinfo {author} {\bibfnamefont {M.}~\bibnamefont {Boareto}},
  \bibinfo {author} {\bibfnamefont {B.}~\bibnamefont {Huang}}, \bibinfo
  {author} {\bibfnamefont {D.}~\bibnamefont {Jia}}, \bibinfo {author}
  {\bibfnamefont {M.}~\bibnamefont {Lu}}, \bibinfo {author} {\bibfnamefont
  {E.}~\bibnamefont {Ben-Jacob}}, \bibinfo {author} {\bibfnamefont {J.~N.}\
  \bibnamefont {Onuchic}}, \ and\ \bibinfo {author} {\bibfnamefont
  {H.}~\bibnamefont {Levine}},\ }\href {\doibase 10.3389/fonc.2015.00155}
  {\bibfield  {journal} {\bibinfo  {journal} {Front. Oncol.}\ }\textbf
  {\bibinfo {volume} {5}},\ \bibinfo {pages} {155} (\bibinfo {year}
  {2015})}\BibitemShut {NoStop}%
\bibitem [{\citenamefont {McMillen}\ \emph {et~al.}(2016)\citenamefont
  {McMillen}, \citenamefont {Chatti}, \citenamefont {J{\"u}lich},\ and\
  \citenamefont {Holley}}]{dev:McMillenCurrBio2016}%
  \BibitemOpen
  \bibfield  {author} {\bibinfo {author} {\bibfnamefont {P.}~\bibnamefont
  {McMillen}}, \bibinfo {author} {\bibfnamefont {V.}~\bibnamefont {Chatti}},
  \bibinfo {author} {\bibfnamefont {D.}~\bibnamefont {J{\"u}lich}}, \ and\
  \bibinfo {author} {\bibfnamefont {S.~A.}\ \bibnamefont {Holley}},\ }\href
  {\doibase 10.1016/j.cub.2015.12.055} {\bibfield  {journal} {\bibinfo
  {journal} {Curr. Biol.}\ }\textbf {\bibinfo {volume} {26}},\ \bibinfo {pages}
  {542 } (\bibinfo {year} {2016})}\BibitemShut {NoStop}%
\bibitem [{\citenamefont {Oswald}\ \emph {et~al.}(2017)\citenamefont {Oswald},
  \citenamefont {Grosser}, \citenamefont {Smith},\ and\ \citenamefont
  {K{\"a}s}}]{cancer:OswaldJPhysDAppPhys2017}%
  \BibitemOpen
  \bibfield  {author} {\bibinfo {author} {\bibfnamefont {L.}~\bibnamefont
  {Oswald}}, \bibinfo {author} {\bibfnamefont {S.}~\bibnamefont {Grosser}},
  \bibinfo {author} {\bibfnamefont {D.~M.}\ \bibnamefont {Smith}}, \ and\
  \bibinfo {author} {\bibfnamefont {J.~A.}\ \bibnamefont {K{\"a}s}},\ }\href
  {\doibase 10.1088/1361-6463/aa8e83} {\bibfield  {journal} {\bibinfo
  {journal} {J. Phys. D Appl. Phys.}\ }\textbf {\bibinfo {volume} {50}},\
  \bibinfo {pages} {483001} (\bibinfo {year} {2017})}\BibitemShut {NoStop}%
\bibitem [{\citenamefont {Mongera}\ \emph {et~al.}(2018)\citenamefont
  {Mongera}, \citenamefont {Rowghanian}, \citenamefont {Gustafson},
  \citenamefont {Shelton}, \citenamefont {Kealhofer}, \citenamefont {Carn},
  \citenamefont {Serwane}, \citenamefont {Lucio}, \citenamefont {Giammona},\
  and\ \citenamefont {Camp{\`a}s}}]{dev:MongeraNature2018}%
  \BibitemOpen
  \bibfield  {author} {\bibinfo {author} {\bibfnamefont {A.}~\bibnamefont
  {Mongera}}, \bibinfo {author} {\bibfnamefont {P.}~\bibnamefont {Rowghanian}},
  \bibinfo {author} {\bibfnamefont {H.~J.}\ \bibnamefont {Gustafson}}, \bibinfo
  {author} {\bibfnamefont {E.}~\bibnamefont {Shelton}}, \bibinfo {author}
  {\bibfnamefont {D.~A.}\ \bibnamefont {Kealhofer}}, \bibinfo {author}
  {\bibfnamefont {E.~K.}\ \bibnamefont {Carn}}, \bibinfo {author}
  {\bibfnamefont {F.}~\bibnamefont {Serwane}}, \bibinfo {author} {\bibfnamefont
  {A.~A.}\ \bibnamefont {Lucio}}, \bibinfo {author} {\bibfnamefont
  {J.}~\bibnamefont {Giammona}}, \ and\ \bibinfo {author} {\bibfnamefont
  {O.}~\bibnamefont {Camp{\`a}s}},\ }\href {\doibase 10.1038/s41586-018-0479-2}
  {\bibfield  {journal} {\bibinfo  {journal} {Nature}\ }\textbf {\bibinfo
  {volume} {561}},\ \bibinfo {pages} {401} (\bibinfo {year}
  {2018})}\BibitemShut {NoStop}%
\bibitem [{\citenamefont {Ilina}\ \emph {et~al.}(2020)\citenamefont {Ilina},
  \citenamefont {Gritsenko}, \citenamefont {Syga}, \citenamefont {Lippoldt},
  \citenamefont {La~Porta}, \citenamefont {Chepizhko}, \citenamefont {Grosser},
  \citenamefont {Vullings}, \citenamefont {Bakker}, \citenamefont
  {Starru{\ss}}, \citenamefont {Bult}, \citenamefont {Zapperi}, \citenamefont
  {K{\"a}s}, \citenamefont {Deutsch},\ and\ \citenamefont
  {Friedl}}]{cancer:IlinaNatCellBio2020}%
  \BibitemOpen
  \bibfield  {author} {\bibinfo {author} {\bibfnamefont {O.}~\bibnamefont
  {Ilina}}, \bibinfo {author} {\bibfnamefont {P.~G.}\ \bibnamefont
  {Gritsenko}}, \bibinfo {author} {\bibfnamefont {S.}~\bibnamefont {Syga}},
  \bibinfo {author} {\bibfnamefont {J.}~\bibnamefont {Lippoldt}}, \bibinfo
  {author} {\bibfnamefont {C.~A.~M.}\ \bibnamefont {La~Porta}}, \bibinfo
  {author} {\bibfnamefont {O.}~\bibnamefont {Chepizhko}}, \bibinfo {author}
  {\bibfnamefont {S.}~\bibnamefont {Grosser}}, \bibinfo {author} {\bibfnamefont
  {M.}~\bibnamefont {Vullings}}, \bibinfo {author} {\bibfnamefont {G.-J.}\
  \bibnamefont {Bakker}}, \bibinfo {author} {\bibfnamefont {J.}~\bibnamefont
  {Starru{\ss}}}, \bibinfo {author} {\bibfnamefont {P.}~\bibnamefont {Bult}},
  \bibinfo {author} {\bibfnamefont {S.}~\bibnamefont {Zapperi}}, \bibinfo
  {author} {\bibfnamefont {J.~A.}\ \bibnamefont {K{\"a}s}}, \bibinfo {author}
  {\bibfnamefont {A.}~\bibnamefont {Deutsch}}, \ and\ \bibinfo {author}
  {\bibfnamefont {P.}~\bibnamefont {Friedl}},\ }\href {\doibase
  10.1038/s41556-020-0552-6} {\bibfield  {journal} {\bibinfo  {journal} {Nat.
  Cell Biol.}\ }\textbf {\bibinfo {volume} {22}},\ \bibinfo {pages} {1103}
  (\bibinfo {year} {2020})}\BibitemShut {NoStop}%
\bibitem [{\citenamefont {Kim}\ \emph {et~al.}(2021)\citenamefont {Kim},
  \citenamefont {Pochitaloff}, \citenamefont {Stooke-Vaughan},\ and\
  \citenamefont {Camp{\`a}s}}]{dev:KimBioRXiv2020}%
  \BibitemOpen
  \bibfield  {author} {\bibinfo {author} {\bibfnamefont {S.}~\bibnamefont
  {Kim}}, \bibinfo {author} {\bibfnamefont {M.}~\bibnamefont {Pochitaloff}},
  \bibinfo {author} {\bibfnamefont {G.~A.}\ \bibnamefont {Stooke-Vaughan}}, \
  and\ \bibinfo {author} {\bibfnamefont {O.}~\bibnamefont {Camp{\`a}s}},\
  }\href {\doibase 10.1038/s41567-021-01215-1} {\bibfield  {journal} {\bibinfo
  {journal} {Nature Physics}\ } (\bibinfo {year} {2021}),\
  10.1038/s41567-021-01215-1}\BibitemShut {NoStop}%
\bibitem [{\citenamefont {Wuyts}\ \emph {et~al.}(2010)\citenamefont {Wuyts},
  \citenamefont {Palauqui}, \citenamefont {Conejero}, \citenamefont {Verdeil},
  \citenamefont {Granier},\ and\ \citenamefont
  {Massonnet}}]{plants:WuytsPlantMethods2010}%
  \BibitemOpen
  \bibfield  {author} {\bibinfo {author} {\bibfnamefont {N.}~\bibnamefont
  {Wuyts}}, \bibinfo {author} {\bibfnamefont {J.-C.}\ \bibnamefont {Palauqui}},
  \bibinfo {author} {\bibfnamefont {G.}~\bibnamefont {Conejero}}, \bibinfo
  {author} {\bibfnamefont {J.-L.}\ \bibnamefont {Verdeil}}, \bibinfo {author}
  {\bibfnamefont {C.}~\bibnamefont {Granier}}, \ and\ \bibinfo {author}
  {\bibfnamefont {C.}~\bibnamefont {Massonnet}},\ }\href {\doibase
  10.1186/1746-4811-6-17} {\bibfield  {journal} {\bibinfo  {journal} {Plant
  Methods}\ }\textbf {\bibinfo {volume} {6}},\ \bibinfo {pages} {17} (\bibinfo
  {year} {2010})}\BibitemShut {NoStop}%
\bibitem [{\citenamefont {Kalve}\ \emph {et~al.}(2014)\citenamefont {Kalve},
  \citenamefont {Fotschki}, \citenamefont {Beeckman}, \citenamefont
  {Vissenberg},\ and\ \citenamefont {Beemster}}]{plants:KalveJExpBot2014}%
  \BibitemOpen
  \bibfield  {author} {\bibinfo {author} {\bibfnamefont {S.}~\bibnamefont
  {Kalve}}, \bibinfo {author} {\bibfnamefont {J.}~\bibnamefont {Fotschki}},
  \bibinfo {author} {\bibfnamefont {T.}~\bibnamefont {Beeckman}}, \bibinfo
  {author} {\bibfnamefont {K.}~\bibnamefont {Vissenberg}}, \ and\ \bibinfo
  {author} {\bibfnamefont {G.~T.~S.}\ \bibnamefont {Beemster}},\ }\href
  {\doibase 10.1093/jxb/eru358} {\bibfield  {journal} {\bibinfo  {journal} {J.
  Exp. Bot.}\ }\textbf {\bibinfo {volume} {65}},\ \bibinfo {pages} {6385}
  (\bibinfo {year} {2014})}\BibitemShut {NoStop}%
\bibitem [{\citenamefont {Sapala}\ \emph {et~al.}(2018)\citenamefont {Sapala},
  \citenamefont {Runions}, \citenamefont {Routier-Kierzkowska}, \citenamefont
  {Das~Gupta}, \citenamefont {Hong}, \citenamefont {Hofhuis}, \citenamefont
  {Verger}, \citenamefont {Mosca}, \citenamefont {Li}, \citenamefont {Hay},
  \citenamefont {Hamant}, \citenamefont {Roeder}, \citenamefont {Tsiantis},
  \citenamefont {Prusinkiewicz},\ and\ \citenamefont
  {Smith}}]{plants:SapalaELife2018}%
  \BibitemOpen
  \bibfield  {author} {\bibinfo {author} {\bibfnamefont {A.}~\bibnamefont
  {Sapala}}, \bibinfo {author} {\bibfnamefont {A.}~\bibnamefont {Runions}},
  \bibinfo {author} {\bibfnamefont {A.-L.}\ \bibnamefont
  {Routier-Kierzkowska}}, \bibinfo {author} {\bibfnamefont {M.}~\bibnamefont
  {Das~Gupta}}, \bibinfo {author} {\bibfnamefont {L.}~\bibnamefont {Hong}},
  \bibinfo {author} {\bibfnamefont {H.}~\bibnamefont {Hofhuis}}, \bibinfo
  {author} {\bibfnamefont {S.}~\bibnamefont {Verger}}, \bibinfo {author}
  {\bibfnamefont {G.}~\bibnamefont {Mosca}}, \bibinfo {author} {\bibfnamefont
  {C.-B.}\ \bibnamefont {Li}}, \bibinfo {author} {\bibfnamefont
  {A.}~\bibnamefont {Hay}}, \bibinfo {author} {\bibfnamefont {O.}~\bibnamefont
  {Hamant}}, \bibinfo {author} {\bibfnamefont {A.~H.}\ \bibnamefont {Roeder}},
  \bibinfo {author} {\bibfnamefont {M.}~\bibnamefont {Tsiantis}}, \bibinfo
  {author} {\bibfnamefont {P.}~\bibnamefont {Prusinkiewicz}}, \ and\ \bibinfo
  {author} {\bibfnamefont {R.~S.}\ \bibnamefont {Smith}},\ }\href {\doibase
  10.7554/eLife.32794} {\bibfield  {journal} {\bibinfo  {journal} {eLife}\
  }\textbf {\bibinfo {volume} {7}},\ \bibinfo {pages} {e32794} (\bibinfo {year}
  {2018})}\BibitemShut {NoStop}%
\bibitem [{\citenamefont {Martinez}\ \emph {et~al.}(2018)\citenamefont
  {Martinez}, \citenamefont {Allsman}, \citenamefont {Brakke}, \citenamefont
  {Hoyt}, \citenamefont {Hayes}, \citenamefont {Liang}, \citenamefont {Neher},
  \citenamefont {Rui}, \citenamefont {Roberts}, \citenamefont {Moradifam},
  \citenamefont {Goldstein}, \citenamefont {Anderson},\ and\ \citenamefont
  {Rasmussen}}]{plants:MartinezThePlantCell2018}%
  \BibitemOpen
  \bibfield  {author} {\bibinfo {author} {\bibfnamefont {P.}~\bibnamefont
  {Martinez}}, \bibinfo {author} {\bibfnamefont {L.~A.}\ \bibnamefont
  {Allsman}}, \bibinfo {author} {\bibfnamefont {K.~A.}\ \bibnamefont {Brakke}},
  \bibinfo {author} {\bibfnamefont {C.}~\bibnamefont {Hoyt}}, \bibinfo {author}
  {\bibfnamefont {J.}~\bibnamefont {Hayes}}, \bibinfo {author} {\bibfnamefont
  {H.}~\bibnamefont {Liang}}, \bibinfo {author} {\bibfnamefont
  {W.}~\bibnamefont {Neher}}, \bibinfo {author} {\bibfnamefont
  {Y.}~\bibnamefont {Rui}}, \bibinfo {author} {\bibfnamefont {A.~M.}\
  \bibnamefont {Roberts}}, \bibinfo {author} {\bibfnamefont {A.}~\bibnamefont
  {Moradifam}}, \bibinfo {author} {\bibfnamefont {B.}~\bibnamefont
  {Goldstein}}, \bibinfo {author} {\bibfnamefont {C.~T.}\ \bibnamefont
  {Anderson}}, \ and\ \bibinfo {author} {\bibfnamefont {C.~G.}\ \bibnamefont
  {Rasmussen}},\ }\href {\doibase 10.1105/tpc.18.00401} {\bibfield  {journal}
  {\bibinfo  {journal} {Plant Cell}\ }\textbf {\bibinfo {volume} {30}},\
  \bibinfo {pages} {2255} (\bibinfo {year} {2018})}\BibitemShut {NoStop}%
\bibitem [{\citenamefont {Borsuk}\ \emph {et~al.}(2019)\citenamefont {Borsuk},
  \citenamefont {Roddy}, \citenamefont {Th{\'e}roux-Rancourt},\ and\
  \citenamefont {Brodersen}}]{plants:BorsukBioRxiv2019}%
  \BibitemOpen
  \bibfield  {author} {\bibinfo {author} {\bibfnamefont {A.~M.}\ \bibnamefont
  {Borsuk}}, \bibinfo {author} {\bibfnamefont {A.~B.}\ \bibnamefont {Roddy}},
  \bibinfo {author} {\bibfnamefont {G.}~\bibnamefont {Th{\'e}roux-Rancourt}}, \
  and\ \bibinfo {author} {\bibfnamefont {C.~R.}\ \bibnamefont {Brodersen}},\
  }\href {\doibase 10.1101/852459} {\bibfield  {journal} {\bibinfo  {journal}
  {bioRxiv}\ } (\bibinfo {year} {2019}),\ 10.1101/852459}\BibitemShut {NoStop}%
\bibitem [{\citenamefont {O'Hern}\ \emph {et~al.}(2003)\citenamefont {O'Hern},
  \citenamefont {Silbert}, \citenamefont {Liu},\ and\ \citenamefont
  {Nagel}}]{jamming:OHernPRE2003}%
  \BibitemOpen
  \bibfield  {author} {\bibinfo {author} {\bibfnamefont {C.~S.}\ \bibnamefont
  {O'Hern}}, \bibinfo {author} {\bibfnamefont {L.~E.}\ \bibnamefont {Silbert}},
  \bibinfo {author} {\bibfnamefont {A.~J.}\ \bibnamefont {Liu}}, \ and\
  \bibinfo {author} {\bibfnamefont {S.~R.}\ \bibnamefont {Nagel}},\ }\href
  {\doibase 10.1103/PhysRevE.68.011306} {\bibfield  {journal} {\bibinfo
  {journal} {Phys. Rev. E}\ }\textbf {\bibinfo {volume} {68}},\ \bibinfo
  {pages} {011306} (\bibinfo {year} {2003})}\BibitemShut {NoStop}%
\bibitem [{\citenamefont {Tkachenko}\ and\ \citenamefont
  {Witten}(1999)}]{contacts:TkachenkoPRE1999}%
  \BibitemOpen
  \bibfield  {author} {\bibinfo {author} {\bibfnamefont {A.~V.}\ \bibnamefont
  {Tkachenko}}\ and\ \bibinfo {author} {\bibfnamefont {T.~A.}\ \bibnamefont
  {Witten}},\ }\href {\doibase 10.1103/PhysRevE.60.687} {\bibfield  {journal}
  {\bibinfo  {journal} {Phys. Rev. E}\ }\textbf {\bibinfo {volume} {60}},\
  \bibinfo {pages} {687} (\bibinfo {year} {1999})}\BibitemShut {NoStop}%
\bibitem [{\citenamefont {Pellegrino}\ and\ \citenamefont
  {Calladine}(1986)}]{frames:PellegrinoIntJSolidsStructs}%
  \BibitemOpen
  \bibfield  {author} {\bibinfo {author} {\bibfnamefont {S.}~\bibnamefont
  {Pellegrino}}\ and\ \bibinfo {author} {\bibfnamefont {C.}~\bibnamefont
  {Calladine}},\ }\href {\doibase https://doi.org/10.1016/0020-7683(86)90014-4}
  {\bibfield  {journal} {\bibinfo  {journal} {International Journal of Solids
  and Structures}\ }\textbf {\bibinfo {volume} {22}},\ \bibinfo {pages} {409}
  (\bibinfo {year} {1986})}\BibitemShut {NoStop}%
\bibitem [{\citenamefont {Schreck}\ \emph
  {et~al.}(2011{\natexlab{a}})\citenamefont {Schreck}, \citenamefont
  {Bertrand}, \citenamefont {O'Hern},\ and\ \citenamefont
  {Shattuck}}]{jamming:SchreckPRL2011}%
  \BibitemOpen
  \bibfield  {author} {\bibinfo {author} {\bibfnamefont {C.~F.}\ \bibnamefont
  {Schreck}}, \bibinfo {author} {\bibfnamefont {T.}~\bibnamefont {Bertrand}},
  \bibinfo {author} {\bibfnamefont {C.~S.}\ \bibnamefont {O'Hern}}, \ and\
  \bibinfo {author} {\bibfnamefont {M.~D.}\ \bibnamefont {Shattuck}},\ }\href
  {\doibase 10.1103/PhysRevLett.107.078301} {\bibfield  {journal} {\bibinfo
  {journal} {Phys. Rev. Lett.}\ }\textbf {\bibinfo {volume} {107}},\ \bibinfo
  {pages} {078301} (\bibinfo {year} {2011}{\natexlab{a}})}\BibitemShut
  {NoStop}%
\bibitem [{\citenamefont {Goodrich}\ \emph {et~al.}(2014)\citenamefont
  {Goodrich}, \citenamefont {Liu},\ and\ \citenamefont
  {Nagel}}]{softp:GoodrichNatPhys2014}%
  \BibitemOpen
  \bibfield  {author} {\bibinfo {author} {\bibfnamefont {C.~P.}\ \bibnamefont
  {Goodrich}}, \bibinfo {author} {\bibfnamefont {A.~J.}\ \bibnamefont {Liu}}, \
  and\ \bibinfo {author} {\bibfnamefont {S.~R.}\ \bibnamefont {Nagel}},\ }\href
  {\doibase 10.1038/nphys3006} {\bibfield  {journal} {\bibinfo  {journal} {Nat.
  Phys.}\ }\textbf {\bibinfo {volume} {10}},\ \bibinfo {pages} {578} (\bibinfo
  {year} {2014})}\BibitemShut {NoStop}%
\bibitem [{\citenamefont {Mailman}\ \emph {et~al.}(2009)\citenamefont
  {Mailman}, \citenamefont {Schreck}, \citenamefont {O'Hern},\ and\
  \citenamefont {Chakraborty}}]{jamming:MailmanPRL2009}%
  \BibitemOpen
  \bibfield  {author} {\bibinfo {author} {\bibfnamefont {M.}~\bibnamefont
  {Mailman}}, \bibinfo {author} {\bibfnamefont {C.~F.}\ \bibnamefont
  {Schreck}}, \bibinfo {author} {\bibfnamefont {C.~S.}\ \bibnamefont {O'Hern}},
  \ and\ \bibinfo {author} {\bibfnamefont {B.}~\bibnamefont {Chakraborty}},\
  }\href {\doibase 10.1103/PhysRevLett.102.255501} {\bibfield  {journal}
  {\bibinfo  {journal} {Phys. Rev. Lett.}\ }\textbf {\bibinfo {volume} {102}},\
  \bibinfo {pages} {255501} (\bibinfo {year} {2009})}\BibitemShut {NoStop}%
\bibitem [{\citenamefont {Donev}\ \emph {et~al.}(2007)\citenamefont {Donev},
  \citenamefont {Connelly}, \citenamefont {Stillinger},\ and\ \citenamefont
  {Torquato}}]{ellipse:DonevPRE2007}%
  \BibitemOpen
  \bibfield  {author} {\bibinfo {author} {\bibfnamefont {A.}~\bibnamefont
  {Donev}}, \bibinfo {author} {\bibfnamefont {R.}~\bibnamefont {Connelly}},
  \bibinfo {author} {\bibfnamefont {F.~H.}\ \bibnamefont {Stillinger}}, \ and\
  \bibinfo {author} {\bibfnamefont {S.}~\bibnamefont {Torquato}},\ }\href
  {\doibase 10.1103/PhysRevE.75.051304} {\bibfield  {journal} {\bibinfo
  {journal} {Phys. Rev. E}\ }\textbf {\bibinfo {volume} {75}},\ \bibinfo
  {pages} {051304} (\bibinfo {year} {2007})}\BibitemShut {NoStop}%
\bibitem [{\citenamefont {Schreck}\ \emph {et~al.}(2012)\citenamefont
  {Schreck}, \citenamefont {Mailman}, \citenamefont {Chakraborty},\ and\
  \citenamefont {O'Hern}}]{jamming:SchreckPRE2012}%
  \BibitemOpen
  \bibfield  {author} {\bibinfo {author} {\bibfnamefont {C.~F.}\ \bibnamefont
  {Schreck}}, \bibinfo {author} {\bibfnamefont {M.}~\bibnamefont {Mailman}},
  \bibinfo {author} {\bibfnamefont {B.}~\bibnamefont {Chakraborty}}, \ and\
  \bibinfo {author} {\bibfnamefont {C.~S.}\ \bibnamefont {O'Hern}},\ }\href
  {\doibase 10.1103/PhysRevE.85.061305} {\bibfield  {journal} {\bibinfo
  {journal} {Phys. Rev. E}\ }\textbf {\bibinfo {volume} {85}},\ \bibinfo
  {pages} {061305} (\bibinfo {year} {2012})}\BibitemShut {NoStop}%
\bibitem [{\citenamefont {VanderWerf}\ \emph {et~al.}(2018)\citenamefont
  {VanderWerf}, \citenamefont {Jin}, \citenamefont {Shattuck},\ and\
  \citenamefont {O'Hern}}]{jamming:VanderWerfPRE2018}%
  \BibitemOpen
  \bibfield  {author} {\bibinfo {author} {\bibfnamefont {K.}~\bibnamefont
  {VanderWerf}}, \bibinfo {author} {\bibfnamefont {W.}~\bibnamefont {Jin}},
  \bibinfo {author} {\bibfnamefont {M.~D.}\ \bibnamefont {Shattuck}}, \ and\
  \bibinfo {author} {\bibfnamefont {C.~S.}\ \bibnamefont {O'Hern}},\ }\href
  {\doibase 10.1103/PhysRevE.97.012909} {\bibfield  {journal} {\bibinfo
  {journal} {Phys. Rev. E}\ }\textbf {\bibinfo {volume} {97}},\ \bibinfo
  {pages} {012909} (\bibinfo {year} {2018})}\BibitemShut {NoStop}%
\bibitem [{\citenamefont {Yuan}\ \emph {et~al.}(2019)\citenamefont {Yuan},
  \citenamefont {VanderWerf}, \citenamefont {Shattuck},\ and\ \citenamefont
  {O'Hern}}]{jamming:YuanSM2019}%
  \BibitemOpen
  \bibfield  {author} {\bibinfo {author} {\bibfnamefont {Y.}~\bibnamefont
  {Yuan}}, \bibinfo {author} {\bibfnamefont {K.}~\bibnamefont {VanderWerf}},
  \bibinfo {author} {\bibfnamefont {M.~D.}\ \bibnamefont {Shattuck}}, \ and\
  \bibinfo {author} {\bibfnamefont {C.~S.}\ \bibnamefont {O'Hern}},\ }\href
  {\doibase 10.1039/C9SM01932D} {\bibfield  {journal} {\bibinfo  {journal}
  {Soft Matter}\ }\textbf {\bibinfo {volume} {15}},\ \bibinfo {pages} {9751}
  (\bibinfo {year} {2019})}\BibitemShut {NoStop}%
\bibitem [{\citenamefont {Brito}\ \emph {et~al.}(2018)\citenamefont {Brito},
  \citenamefont {Ikeda}, \citenamefont {Urbani}, \citenamefont {Wyart},\ and\
  \citenamefont {Zamponi}}]{nonspherical:BritoPNAS2018}%
  \BibitemOpen
  \bibfield  {author} {\bibinfo {author} {\bibfnamefont {C.}~\bibnamefont
  {Brito}}, \bibinfo {author} {\bibfnamefont {H.}~\bibnamefont {Ikeda}},
  \bibinfo {author} {\bibfnamefont {P.}~\bibnamefont {Urbani}}, \bibinfo
  {author} {\bibfnamefont {M.}~\bibnamefont {Wyart}}, \ and\ \bibinfo {author}
  {\bibfnamefont {F.}~\bibnamefont {Zamponi}},\ }\href {\doibase
  10.1073/pnas.1812457115} {\bibfield  {journal} {\bibinfo  {journal} {Proc.
  Natl. Acad. Sci. USA}\ }\textbf {\bibinfo {volume} {115}},\ \bibinfo {pages}
  {11736} (\bibinfo {year} {2018})}\BibitemShut {NoStop}%
\bibitem [{\citenamefont {Shen}\ \emph {et~al.}(2012)\citenamefont {Shen},
  \citenamefont {Schreck}, \citenamefont {Chakraborty}, \citenamefont {Freed},\
  and\ \citenamefont {O'Hern}}]{jamming:ShenPRE2013}%
  \BibitemOpen
  \bibfield  {author} {\bibinfo {author} {\bibfnamefont {T.}~\bibnamefont
  {Shen}}, \bibinfo {author} {\bibfnamefont {C.}~\bibnamefont {Schreck}},
  \bibinfo {author} {\bibfnamefont {B.}~\bibnamefont {Chakraborty}}, \bibinfo
  {author} {\bibfnamefont {D.~E.}\ \bibnamefont {Freed}}, \ and\ \bibinfo
  {author} {\bibfnamefont {C.~S.}\ \bibnamefont {O'Hern}},\ }\href {\doibase
  10.1103/PhysRevE.86.041303} {\bibfield  {journal} {\bibinfo  {journal} {Phys.
  Rev. E}\ }\textbf {\bibinfo {volume} {86}},\ \bibinfo {pages} {041303}
  (\bibinfo {year} {2012})}\BibitemShut {NoStop}%
\bibitem [{\citenamefont {Damavandi}\ \emph {et~al.}(2021)\citenamefont
  {Damavandi}, \citenamefont {Hagh}, \citenamefont {Santangelo},\ and\
  \citenamefont {Manning}}]{rigidity:DamavindiArXiv2021}%
  \BibitemOpen
  \bibfield  {author} {\bibinfo {author} {\bibfnamefont {O.~K.}\ \bibnamefont
  {Damavandi}}, \bibinfo {author} {\bibfnamefont {V.~F.}\ \bibnamefont {Hagh}},
  \bibinfo {author} {\bibfnamefont {C.~D.}\ \bibnamefont {Santangelo}}, \ and\
  \bibinfo {author} {\bibfnamefont {M.~L.}\ \bibnamefont {Manning}},\
  }\href@noop {} {\  (\bibinfo {year} {2021})},\ \Eprint
  {http://arxiv.org/abs/2102.11310} {arXiv:2102.11310} \BibitemShut {NoStop}%
\bibitem [{\citenamefont {Bi}\ \emph {et~al.}(2015)\citenamefont {Bi},
  \citenamefont {Lopez}, \citenamefont {Schwarz},\ and\ \citenamefont
  {Manning}}]{vertex:BiNatPhys2015}%
  \BibitemOpen
  \bibfield  {author} {\bibinfo {author} {\bibfnamefont {D.}~\bibnamefont
  {Bi}}, \bibinfo {author} {\bibfnamefont {J.~H.}\ \bibnamefont {Lopez}},
  \bibinfo {author} {\bibfnamefont {J.~M.}\ \bibnamefont {Schwarz}}, \ and\
  \bibinfo {author} {\bibfnamefont {M.~L.}\ \bibnamefont {Manning}},\ }\href
  {\doibase 10.1038/nphys3471} {\bibfield  {journal} {\bibinfo  {journal} {Nat.
  Phys.}\ }\textbf {\bibinfo {volume} {11}},\ \bibinfo {pages} {1074} (\bibinfo
  {year} {2015})}\BibitemShut {NoStop}%
\bibitem [{\citenamefont {Yan}\ and\ \citenamefont
  {Bi}(2019)}]{vertex:LePRX2019}%
  \BibitemOpen
  \bibfield  {author} {\bibinfo {author} {\bibfnamefont {L.}~\bibnamefont
  {Yan}}\ and\ \bibinfo {author} {\bibfnamefont {D.}~\bibnamefont {Bi}},\
  }\href {\doibase 10.1103/PhysRevX.9.011029} {\bibfield  {journal} {\bibinfo
  {journal} {Phys. Rev. X}\ }\textbf {\bibinfo {volume} {9}},\ \bibinfo {pages}
  {011029} (\bibinfo {year} {2019})}\BibitemShut {NoStop}%
\bibitem [{\citenamefont {Boromand}\ \emph {et~al.}(2018)\citenamefont
  {Boromand}, \citenamefont {Signoriello}, \citenamefont {Ye}, \citenamefont
  {O'Hern},\ and\ \citenamefont {Shattuck}}]{jamming:BoromandPRL2018}%
  \BibitemOpen
  \bibfield  {author} {\bibinfo {author} {\bibfnamefont {A.}~\bibnamefont
  {Boromand}}, \bibinfo {author} {\bibfnamefont {A.}~\bibnamefont
  {Signoriello}}, \bibinfo {author} {\bibfnamefont {F.}~\bibnamefont {Ye}},
  \bibinfo {author} {\bibfnamefont {C.~S.}\ \bibnamefont {O'Hern}}, \ and\
  \bibinfo {author} {\bibfnamefont {M.~D.}\ \bibnamefont {Shattuck}},\ }\href
  {\doibase 10.1103/PhysRevLett.121.248003} {\bibfield  {journal} {\bibinfo
  {journal} {Phys. Rev. Lett.}\ }\textbf {\bibinfo {volume} {121}},\ \bibinfo
  {pages} {248003} (\bibinfo {year} {2018})}\BibitemShut {NoStop}%
\bibitem [{\citenamefont {Zhang}\ \emph {et~al.}(2014)\citenamefont {Zhang},
  \citenamefont {Smith}, \citenamefont {Wang}, \citenamefont {Liu},
  \citenamefont {Schroers}, \citenamefont {Shattuck},\ and\ \citenamefont
  {O'Hern}}]{jamming:ZhangPRE2014}%
  \BibitemOpen
  \bibfield  {author} {\bibinfo {author} {\bibfnamefont {K.}~\bibnamefont
  {Zhang}}, \bibinfo {author} {\bibfnamefont {W.~W.}\ \bibnamefont {Smith}},
  \bibinfo {author} {\bibfnamefont {M.}~\bibnamefont {Wang}}, \bibinfo {author}
  {\bibfnamefont {Y.}~\bibnamefont {Liu}}, \bibinfo {author} {\bibfnamefont
  {J.}~\bibnamefont {Schroers}}, \bibinfo {author} {\bibfnamefont {M.~D.}\
  \bibnamefont {Shattuck}}, \ and\ \bibinfo {author} {\bibfnamefont {C.~S.}\
  \bibnamefont {O'Hern}},\ }\href {\doibase 10.1103/PhysRevE.90.032311}
  {\bibfield  {journal} {\bibinfo  {journal} {Phys. Rev. E}\ }\textbf {\bibinfo
  {volume} {90}},\ \bibinfo {pages} {032311} (\bibinfo {year}
  {2014})}\BibitemShut {NoStop}%
\bibitem [{\citenamefont {Kane}\ and\ \citenamefont
  {Lubensky}(2014)}]{metamat:LubenskyNatPhys2014}%
  \BibitemOpen
  \bibfield  {author} {\bibinfo {author} {\bibfnamefont {C.~L.}\ \bibnamefont
  {Kane}}\ and\ \bibinfo {author} {\bibfnamefont {T.~C.}\ \bibnamefont
  {Lubensky}},\ }\href {\doibase 10.1038/nphys2835} {\bibfield  {journal}
  {\bibinfo  {journal} {Nature Physics}\ }\textbf {\bibinfo {volume} {10}},\
  \bibinfo {pages} {39} (\bibinfo {year} {2014})}\BibitemShut {NoStop}%
\bibitem [{\citenamefont {Chen}\ \emph {et~al.}(2014)\citenamefont {Chen},
  \citenamefont {Upadhyaya},\ and\ \citenamefont
  {Vitelli}}]{metamat:ChenPNAS2014}%
  \BibitemOpen
  \bibfield  {author} {\bibinfo {author} {\bibfnamefont {B.~G.-g.}\
  \bibnamefont {Chen}}, \bibinfo {author} {\bibfnamefont {N.}~\bibnamefont
  {Upadhyaya}}, \ and\ \bibinfo {author} {\bibfnamefont {V.}~\bibnamefont
  {Vitelli}},\ }\href {\doibase 10.1073/pnas.1405969111} {\bibfield  {journal}
  {\bibinfo  {journal} {Proceedings of the National Academy of Sciences}\
  }\textbf {\bibinfo {volume} {111}},\ \bibinfo {pages} {13004} (\bibinfo
  {year} {2014})}\BibitemShut {NoStop}%
\bibitem [{\citenamefont {Papanikolaou}\ \emph {et~al.}(2013)\citenamefont
  {Papanikolaou}, \citenamefont {O'Hern},\ and\ \citenamefont
  {Shattuck}}]{jamming:PapaPRL2013}%
  \BibitemOpen
  \bibfield  {author} {\bibinfo {author} {\bibfnamefont {S.}~\bibnamefont
  {Papanikolaou}}, \bibinfo {author} {\bibfnamefont {C.~S.}\ \bibnamefont
  {O'Hern}}, \ and\ \bibinfo {author} {\bibfnamefont {M.~D.}\ \bibnamefont
  {Shattuck}},\ }\href {\doibase 10.1103/PhysRevLett.110.198002} {\bibfield
  {journal} {\bibinfo  {journal} {Phys. Rev. Lett.}\ }\textbf {\bibinfo
  {volume} {110}},\ \bibinfo {pages} {198002} (\bibinfo {year}
  {2013})}\BibitemShut {NoStop}%
\bibitem [{\citenamefont {Schreck}\ \emph
  {et~al.}(2011{\natexlab{b}})\citenamefont {Schreck}, \citenamefont {O'Hern},\
  and\ \citenamefont {Silbert}}]{jamming:SchreckPRE2011}%
  \BibitemOpen
  \bibfield  {author} {\bibinfo {author} {\bibfnamefont {C.~F.}\ \bibnamefont
  {Schreck}}, \bibinfo {author} {\bibfnamefont {C.~S.}\ \bibnamefont {O'Hern}},
  \ and\ \bibinfo {author} {\bibfnamefont {L.~E.}\ \bibnamefont {Silbert}},\
  }\href {\doibase 10.1103/PhysRevE.84.011305} {\bibfield  {journal} {\bibinfo
  {journal} {Phys. Rev. E}\ }\textbf {\bibinfo {volume} {84}},\ \bibinfo
  {pages} {011305} (\bibinfo {year} {2011}{\natexlab{b}})}\BibitemShut
  {NoStop}%
\bibitem [{\citenamefont {Tuckman}\ \emph {et~al.}(2020)\citenamefont
  {Tuckman}, \citenamefont {VanderWerf}, \citenamefont {Yuan}, \citenamefont
  {Zhang}, \citenamefont {Zhang}, \citenamefont {Shattuck},\ and\ \citenamefont
  {O’Hern}}]{jamming:TuckmanSM2020}%
  \BibitemOpen
  \bibfield  {author} {\bibinfo {author} {\bibfnamefont {P.~J.}\ \bibnamefont
  {Tuckman}}, \bibinfo {author} {\bibfnamefont {K.}~\bibnamefont {VanderWerf}},
  \bibinfo {author} {\bibfnamefont {Y.}~\bibnamefont {Yuan}}, \bibinfo {author}
  {\bibfnamefont {S.}~\bibnamefont {Zhang}}, \bibinfo {author} {\bibfnamefont
  {J.}~\bibnamefont {Zhang}}, \bibinfo {author} {\bibfnamefont {M.~D.}\
  \bibnamefont {Shattuck}}, \ and\ \bibinfo {author} {\bibfnamefont {C.~S.}\
  \bibnamefont {O’Hern}},\ }\href {\doibase 10.1039/D0SM01137A} {\bibfield
  {journal} {\bibinfo  {journal} {Soft Matter}\ }\textbf {\bibinfo {volume}
  {16}},\ \bibinfo {pages} {9443} (\bibinfo {year} {2020})}\BibitemShut
  {NoStop}%
\bibitem [{\citenamefont {Foglino}\ \emph {et~al.}(2017)\citenamefont
  {Foglino}, \citenamefont {Morozov}, \citenamefont {Henrich},\ and\
  \citenamefont {Marenduzzo}}]{droplets:FoglinoPRL2017}%
  \BibitemOpen
  \bibfield  {author} {\bibinfo {author} {\bibfnamefont {M.}~\bibnamefont
  {Foglino}}, \bibinfo {author} {\bibfnamefont {A.~N.}\ \bibnamefont
  {Morozov}}, \bibinfo {author} {\bibfnamefont {O.}~\bibnamefont {Henrich}}, \
  and\ \bibinfo {author} {\bibfnamefont {D.}~\bibnamefont {Marenduzzo}},\
  }\href {\doibase 10.1103/PhysRevLett.119.208002} {\bibfield  {journal}
  {\bibinfo  {journal} {Phys. Rev. Lett.}\ }\textbf {\bibinfo {volume} {119}},\
  \bibinfo {pages} {208002} (\bibinfo {year} {2017})}\BibitemShut {NoStop}%
\bibitem [{\citenamefont {Hong}\ \emph {et~al.}(2017)\citenamefont {Hong},
  \citenamefont {Kohne}, \citenamefont {Morrell}, \citenamefont {Wang},\ and\
  \citenamefont {Weeks}}]{weeks:HongPRE2017}%
  \BibitemOpen
  \bibfield  {author} {\bibinfo {author} {\bibfnamefont {X.}~\bibnamefont
  {Hong}}, \bibinfo {author} {\bibfnamefont {M.}~\bibnamefont {Kohne}},
  \bibinfo {author} {\bibfnamefont {M.}~\bibnamefont {Morrell}}, \bibinfo
  {author} {\bibfnamefont {H.}~\bibnamefont {Wang}}, \ and\ \bibinfo {author}
  {\bibfnamefont {E.~R.}\ \bibnamefont {Weeks}},\ }\href {\doibase
  10.1103/PhysRevE.96.062605} {\bibfield  {journal} {\bibinfo  {journal} {Phys.
  Rev. E}\ }\textbf {\bibinfo {volume} {96}},\ \bibinfo {pages} {062605}
  (\bibinfo {year} {2017})}\BibitemShut {NoStop}%
\bibitem [{\citenamefont {Golovkova}\ \emph {et~al.}(2020)\citenamefont
  {Golovkova}, \citenamefont {Montel}, \citenamefont {Wandersman},
  \citenamefont {Bertrand}, \citenamefont {Prevost},\ and\ \citenamefont
  {Pontani}}]{emulsions:GolovkovaSM2020}%
  \BibitemOpen
  \bibfield  {author} {\bibinfo {author} {\bibfnamefont {I.}~\bibnamefont
  {Golovkova}}, \bibinfo {author} {\bibfnamefont {L.}~\bibnamefont {Montel}},
  \bibinfo {author} {\bibfnamefont {E.}~\bibnamefont {Wandersman}}, \bibinfo
  {author} {\bibfnamefont {T.}~\bibnamefont {Bertrand}}, \bibinfo {author}
  {\bibfnamefont {A.~M.}\ \bibnamefont {Prevost}}, \ and\ \bibinfo {author}
  {\bibfnamefont {L.-L.}\ \bibnamefont {Pontani}},\ }\href {\doibase
  10.1039/C9SM02343G} {\bibfield  {journal} {\bibinfo  {journal} {Soft Matter}\
  }\textbf {\bibinfo {volume} {16}},\ \bibinfo {pages} {3294} (\bibinfo {year}
  {2020})}\BibitemShut {NoStop}%
\bibitem [{\citenamefont {Allen}\ and\ \citenamefont
  {Tildesley}(2017)}]{sim:AllenOxford2017}%
  \BibitemOpen
  \bibfield  {author} {\bibinfo {author} {\bibfnamefont {M.~P.}\ \bibnamefont
  {Allen}}\ and\ \bibinfo {author} {\bibfnamefont {D.~J.}\ \bibnamefont
  {Tildesley}},\ }\href@noop {} {\emph {\bibinfo {title} {Computer Simulation
  of Liquids}}},\ \bibinfo {edition} {2nd}\ ed.\ (\bibinfo  {publisher} {Oxford
  University Press},\ \bibinfo {year} {2017})\BibitemShut {NoStop}%
\bibitem [{\citenamefont {Goodrich}\ \emph {et~al.}(2012)\citenamefont
  {Goodrich}, \citenamefont {Liu},\ and\ \citenamefont
  {Nagel}}]{softp:GoodrichPRL2012}%
  \BibitemOpen
  \bibfield  {author} {\bibinfo {author} {\bibfnamefont {C.~P.}\ \bibnamefont
  {Goodrich}}, \bibinfo {author} {\bibfnamefont {A.~J.}\ \bibnamefont {Liu}}, \
  and\ \bibinfo {author} {\bibfnamefont {S.~R.}\ \bibnamefont {Nagel}},\ }\href
  {\doibase 10.1103/PhysRevLett.109.095704} {\bibfield  {journal} {\bibinfo
  {journal} {Phys. Rev. Lett.}\ }\textbf {\bibinfo {volume} {109}},\ \bibinfo
  {pages} {095704} (\bibinfo {year} {2012})}\BibitemShut {NoStop}%
\bibitem [{\citenamefont {VanderWerf}\ \emph {et~al.}(2020)\citenamefont
  {VanderWerf}, \citenamefont {Boromand}, \citenamefont {Shattuck},\ and\
  \citenamefont {O'Hern}}]{jamming:VanderWerfPRL2020}%
  \BibitemOpen
  \bibfield  {author} {\bibinfo {author} {\bibfnamefont {K.}~\bibnamefont
  {VanderWerf}}, \bibinfo {author} {\bibfnamefont {A.}~\bibnamefont
  {Boromand}}, \bibinfo {author} {\bibfnamefont {M.~D.}\ \bibnamefont
  {Shattuck}}, \ and\ \bibinfo {author} {\bibfnamefont {C.~S.}\ \bibnamefont
  {O'Hern}},\ }\href {\doibase 10.1103/PhysRevLett.124.038004} {\bibfield
  {journal} {\bibinfo  {journal} {Phys. Rev. Lett.}\ }\textbf {\bibinfo
  {volume} {124}},\ \bibinfo {pages} {038004} (\bibinfo {year}
  {2020})}\BibitemShut {NoStop}%
\bibitem [{\citenamefont {Somfai}\ \emph {et~al.}(2007)\citenamefont {Somfai},
  \citenamefont {van Hecke}, \citenamefont {Ellenbroek}, \citenamefont
  {Shundyak},\ and\ \citenamefont {van Saarloos}}]{frictional:SomfaiPRE2007}%
  \BibitemOpen
  \bibfield  {author} {\bibinfo {author} {\bibfnamefont {E.}~\bibnamefont
  {Somfai}}, \bibinfo {author} {\bibfnamefont {M.}~\bibnamefont {van Hecke}},
  \bibinfo {author} {\bibfnamefont {W.~G.}\ \bibnamefont {Ellenbroek}},
  \bibinfo {author} {\bibfnamefont {K.}~\bibnamefont {Shundyak}}, \ and\
  \bibinfo {author} {\bibfnamefont {W.}~\bibnamefont {van Saarloos}},\ }\href
  {\doibase 10.1103/PhysRevE.75.020301} {\bibfield  {journal} {\bibinfo
  {journal} {Phys. Rev. E}\ }\textbf {\bibinfo {volume} {75}},\ \bibinfo
  {pages} {020301(R)} (\bibinfo {year} {2007})}\BibitemShut {NoStop}%
\bibitem [{\citenamefont {Wang}\ \emph {et~al.}(2021)\citenamefont {Wang},
  \citenamefont {Zhang}, \citenamefont {Tuckman}, \citenamefont {Ouellette},
  \citenamefont {Shattuck},\ and\ \citenamefont
  {O'Hern}}]{jamming:WangPRE2021}%
  \BibitemOpen
  \bibfield  {author} {\bibinfo {author} {\bibfnamefont {P.}~\bibnamefont
  {Wang}}, \bibinfo {author} {\bibfnamefont {S.}~\bibnamefont {Zhang}},
  \bibinfo {author} {\bibfnamefont {P.}~\bibnamefont {Tuckman}}, \bibinfo
  {author} {\bibfnamefont {N.~T.}\ \bibnamefont {Ouellette}}, \bibinfo {author}
  {\bibfnamefont {M.~D.}\ \bibnamefont {Shattuck}}, \ and\ \bibinfo {author}
  {\bibfnamefont {C.~S.}\ \bibnamefont {O'Hern}},\ }\href {\doibase
  10.1103/PhysRevE.103.022902} {\bibfield  {journal} {\bibinfo  {journal}
  {Phys. Rev. E}\ }\textbf {\bibinfo {volume} {103}},\ \bibinfo {pages}
  {022902} (\bibinfo {year} {2021})}\BibitemShut {NoStop}%
\bibitem [{\citenamefont {Ajeti}\ \emph {et~al.}(2019)\citenamefont {Ajeti},
  \citenamefont {Tabatabai}, \citenamefont {Fleszar}, \citenamefont {Staddon},
  \citenamefont {Seara}, \citenamefont {Suarez}, \citenamefont {Yousafzai},
  \citenamefont {Bi}, \citenamefont {Kovar}, \citenamefont {Banerjee},\ and\
  \citenamefont {Murrell}}]{collective:AjetiNatPhys2019}%
  \BibitemOpen
  \bibfield  {author} {\bibinfo {author} {\bibfnamefont {V.}~\bibnamefont
  {Ajeti}}, \bibinfo {author} {\bibfnamefont {A.~P.}\ \bibnamefont
  {Tabatabai}}, \bibinfo {author} {\bibfnamefont {A.~J.}\ \bibnamefont
  {Fleszar}}, \bibinfo {author} {\bibfnamefont {M.~F.}\ \bibnamefont
  {Staddon}}, \bibinfo {author} {\bibfnamefont {D.~S.}\ \bibnamefont {Seara}},
  \bibinfo {author} {\bibfnamefont {C.}~\bibnamefont {Suarez}}, \bibinfo
  {author} {\bibfnamefont {M.~S.}\ \bibnamefont {Yousafzai}}, \bibinfo {author}
  {\bibfnamefont {D.}~\bibnamefont {Bi}}, \bibinfo {author} {\bibfnamefont
  {D.~R.}\ \bibnamefont {Kovar}}, \bibinfo {author} {\bibfnamefont
  {S.}~\bibnamefont {Banerjee}}, \ and\ \bibinfo {author} {\bibfnamefont
  {M.~P.}\ \bibnamefont {Murrell}},\ }\href {\doibase
  10.1038/s41567-019-0485-9} {\bibfield  {journal} {\bibinfo  {journal} {Nat.
  Phys.}\ }\textbf {\bibinfo {volume} {15}},\ \bibinfo {pages} {696} (\bibinfo
  {year} {2019})}\BibitemShut {NoStop}%
\end{thebibliography}%

\end{document}